\documentstyle [aaspptwo]{article}
\def\aa{A\&A}
\def\gsimeq
{\hbox{\raise0.5ex\hbox{$>\lower1.06ex\hbox{$\kern-1.07em{\sim}$}$}}}
\def\lsimeq
{\hbox{\raise0.5ex\hbox{$<\lower1.06ex\hbox{$\kern-1.07em{\sim}$}$}}}
\def\pn{\par\noindent}
\def\ss{\smallskip\pn}
\def\ms{\medskip\pn}
\def\bs{\bigskip\pn}

\begin{document}
\title{{\it ASCA} and {\it ROSAT} X-ray Spectra of High-Redshift Radio-Loud Quasars}
\author{M. Cappi$^{1,\dagger}$, M. Matsuoka$^1$, A. Comastri$^2$, W. Brinkmann$^3$, \\M. Elvis $^4$, 
G.G.C. Palumbo$^{5,6}$ and C. Vignali$^5$}

\affil{$^1$ The Institute of Physical and Chemical Research (RIKEN),
2-1, Hirosawa, Wako, Saitama 351-01, Japan}

\affil{$^2$ Osservatorio Astronomico di Bologna, Via Zamboni 33, I-40126 Bologna, Italy}

\affil{$^3$ Max-Planck-Institut f\"ur Extraterrestrische Physik, D-85748 Garching bei
M\"unchen, FRG}

\affil{$^4$ Harvard-Smithsonian Center for Astrophysics, 60 Garden Street, Cambridge, 
Massachussets 02138, USA}

\affil{$^5$ Dipartimento di Astronomia, Universit\`a di Bologna, Via Zamboni 33,
I-40126 Bologna, Italy}

\affil{$^6$ Istituto per le Tecnologie e Studio Radiazioni Extraterrestri, ITeSRE/CNR,
Via Gobetti 101, I-40129 Bologna, Italy}

\affil{$\dagger$ Present address: Istituto per le Tecnologie e Studio Radiazioni 
Extraterrestri, ITeSRE/CNR, Via Gobetti 101, I-40129 Bologna, Italy - 
E-mail: mcappi@tesre.bo.cnr.it}

\ms
\centerline{To be published on The Astrophysical Journal, April 1, 1997 issue, Vol. 478}


\begin{abstract}


Results are presented on the X-ray properties of 9 high-redshift (1.2 $<$ z $<$ 3.4)
radio-loud quasars (RLQs) observed by {\it ASCA} (10 observations) and {\it ROSAT}
(11 observations, for a subset of 6 quasars). New {\it ASCA}
observations of S5 0014+81 (z = 3.38) and S5 0836+71 (z = 2.17) and {\it ROSAT}
observations of PKS 2126-158 for which results were never presented elsewhere, are 
included.

A simple model consisting of a power law plus cold, uniform absorption
gives acceptable fits to the spectra of all sources. The {\it ASCA} spectra
of the 6 brightest objects show evidence for absorption in excess of the Galactic value at a
$\gg$ 99\% confidence level. Comparison with the {\it ROSAT} data suggests that
absorption has significantly varied ($\Delta N_{\rm H}\ \sim 8$ $\times$
10$^{20}$ cm$^{-2}$) in the case of S5 0836+71, on a time-scale of
$\sim$ 0.8 yr in the quasar frame. For the remaining 5 sources for which
{\it ROSAT} spectra were available, the two instruments gave consistent results
and the data were combined yielding unprecedent spectral coverage (typically
$\sim$ 0.4-40 keV in the quasar frame) for high-z quasars. This
allows to put severe limits on several different descriptions of the
continuum (e.g. broken power law, bremsstrahlung, reflection component).
No Fe K$\alpha$ emission line is detected in any of the {\it ASCA} spectra.
An absorption edge consistent with Fe K$\alpha$ at the quasar redshift
is marginally detected in S5 0014+81.
Possible origins for the observed low energy absorption are discussed. In particular,
contributions from the molecular clouds and dust present in our Galaxy 
(usually disregarded) are carefully considered. In the light of the new results 
for S5 0836+71 and S5 0014+81, absorption intrinsic to the quasars is considered and 
discussed.

The average slope obtained from the 8 {\it ASCA} spectra
in the observed $\sim$ 0.5--10 keV energy band is $<\Gamma_{0.5-10 \rm keV}>$ 
$\simeq$ 1.61 $\pm$ 0.04, with a dispersion $\sigma_{0.5-10 \rm keV}$ 
$\simeq$ 0.10 $\pm$ 0.03.
The average photon index in the observed 2--10 keV band, where the effect of absorption is
negligible, is $<\Gamma_{2-10 \rm keV}>$ $\simeq$ 1.53 $\pm$ 0.05, with a dispersion
$\sigma_{2-10 \rm keV}$ $\lsimeq$ 0.12.
Furthermore, the implications of the present results on the calculations of the
contribution of quasars to the cosmic X- and $\gamma$-ray backgrounds (XRB and GRB) 
are briefly discussed.


\keywords{galaxies: active --- quasars: general --- X-rays: galaxies }

\end{abstract}

\normalsize
\onecolumn

\section{Introduction}

Quasars are the most powerful objects
in the whole Universe. This is especially true in the X-ray band, where luminosities
can reach $\sim$ 10$^{47-48}$ erg s$^{-1}$. However how quasars produce such a
large amount of energy remains a challenging astrophysical problem.
Certainly, because of their extreme
conditions, quasars provide a powerful
test for models of emission mechanisms of active galactic nuclei (AGNs) (Rees 1984).
Quasars show strong continuum emission over the entire electro-magnetic spectrum,
from radio through the X- and even $\gamma$-ray region
(Sanders et al. 1989, Elvis et al. 1994a, Thompson et al. 1995).
Optically selected samples of quasars indicate
that $\sim$ 90\% of them are radio quiet
(RQQs) and $\sim$ 10\% are radio loud (RLQs).

X-ray quasar spectral observations are crucial for two
main reasons: X-rays carry a large amount of the total quasar luminosity.
Secondly, as demonstrated by observations of rapid X-ray variability, X-rays 
originate from the innermost regions of the quasar (Mushotzky, Done \& Pounds 1993).
Most X-ray spectral observations have included mainly low redshift (z $<$ 1)
quasars; the poor energy resolution generally limited the analysis
to a simple parameterization of the spectrum with a single power law.
In the $\sim$ 0.1--4 keV energy range, previous {\it Einstein} IPC and {\it ROSAT}
PSPC observations have shown that RLQs have significantly flatter X-ray spectra
than RQQs (Wilkes \& Elvis 1987, Brunner et al. 1992)
and that, for a given optical luminosity, RLQs are on average $\sim$ 3 times brighter
in X-rays than RQQs (Zamorani et al. 1981).
Studies at higher energies ($\sim$ 2--10 keV) with {\it EXOSAT} and
{\it Ginga} have confirmed the dichotomy, with a clear correlation between spectral index and
radio-loudness (Williams et al. 1992, Lawson et al. 1992). Whether differences
in the observed X-ray properties should be attributed either to intrinsically
different properties of the sources or inclination effects
and/or host galaxy properties is not yet well understood. Also selection effects and/or
complex spectral structures (e.g. soft-excess emission, ionized absorption)
may complicate the correct interpretation of the data (Halpern 1984, Comastri et al. 1992,
Fiore et al. 1993).

At high redshifts ($\gsimeq$ 1), spectral information is almost absent for RQQs,
and scarce for RLQs, since soft X-ray observations (mainly from the {\it ROSAT} PSPC)
have allowed only a poor determination of the spectral slopes for only a small number of
objects, mostly RLQs (Bechtold et al. 1994, Elvis et al. 1994b, hereafter E94).
It is not yet clear whether quasars do exhibit spectral evolution. This is a
fundamental question which has a direct impact on quasar formation models.
A remarkable result has been the discovery made with the PSPC that at least some
of the high-z RLQs have low energy cut-offs possibly due to absorption
along the line of sight (E94).
Preliminary results of {\it ASCA} observations of high-z RLQs have indeed already
confirmed the low energy cut-off in two of them (Serlemitsos et al. 1994) and discovered a
third probable case (Siebert et al. 1996, hereafter S96).
Comparison with two high S/N spectra of two RQQs at z $\sim$ 1 (Nandra et al. 1995)
indicates that RQQs are steeper than RLQ even at z $\gsimeq$ 1. However, the
number of quasars observed so far is too small to draw any reliable conclusion.

This paper presents a comprehensive and uniform study of {\it ASCA} observations
of a sample of 9 RLQs with 1.2 $<$ z $<$ 3.4. Whenever possible, the {\it ASCA} results are
compared and combined to {\it ROSAT} spectra extracted from the
public archive. Extensive search for Fe K emission lines, high-energy
excesses (``hard tails'') and alternative models are presented. The
possible origin of the apparently common excess absorption found in the data
is discussed in the light of two newly discovered RLQs with such feature.
Finally, the impact of these new measurements on the cosmic high energy background
radiation is briefly discussed.

In the following, H$_0$ = 50 km s$^{-1}$ Mpc$^{-1}$ and q$_0$=0 are assumed
throughout.

\section{Observations and Data Analysis}

\subsection{The Sample}

The quasar sample consists of all objects (9) either from PI or archival
{\it ASCA} observations available before 1996, January 1$^{\rm st}$. The purpose
was to analyze data of a reasonably large number of quasars to be able to address
for the first time statistically the X-ray properties of the class.

A total of 10 {\it ASCA} pointed observations were collected, with S5 0014+81
observed twice. For five quasars, 11 {\it ROSAT} PSPC observations were retrieved from
the archive and the source spectra re-analyzed to ensure a uniform and
consistent analysis within the sample.
A considerable part of our analysis
reproduces in part previous work on individual sources (see \S 4).
The present analysis, however, differs from single object studies as:
a) it provides a uniform analysis of the quasar sample, b) it makes use of
the most recent calibrations (particularly important for those observations
performed during the Performance and Verification phase of {\it ASCA}) and
c) it compares on a uniform basis {\it ASCA} and {\it ROSAT}
spectral results. It is worth pointing out that {\it ASCA} results on 
S5 0014+81 (z=3.38) and S5 0836+71 (z=2.17) are new.
The relevant data for the whole sample are given in Table 1.

\subsection{{\it ROSAT} Data Reduction}

A sub-sample of six of the 9 quasars were observed on-axis with the Position
Sensitive Proportional Counter (PSPC) (Pfefferman et al. 1987)  on
board the {\it ROSAT} Observatory (Tr\"umper 1983) between 1991 and 1993.
The relevant data for the observations are listed in Table 2.
As indicated in the Table, two new PSPC observations of PKS 2126-158
are presented, which almost double the total number of counts available
for that source.
The PSPC has an energy bandpass in the range 0.1--2.4 keV with
an energy resolution of $\Delta$E/E $\sim$ 0.5 keV at 1 keV.
Source spectra were extracted from circles of $\sim$ 90''--200'' centered on the
sources, and background spectra were taken from annuli centered on the sources or
from circular regions uncontaminated by nearby sources. Source and background counts
were corrected for telescope vignetting.
Data preparation and analysis were performed using the JAN95 version of the
EXSAS/MIDAS software package (Zimmerman et al. 1993).
Spectral analysis was
performed using the version 8.50 of the XSPEC program
(Arnaud, Haberl \& Tennant 1991).

\subsection{{\it ASCA} Data Reduction}

{\it ASCA} has two gas imaging spectrometers (GIS) and two solid-state
imaging spectrometers (SIS) (Tanaka, Inoue \& Holt 1994). The energy resolution of the
GIS and SIS are $\Delta$E/E $\sim$ 0.15 keV and $\sim$ 0.05 keV at 1 keV,
respectively, which
is about 3 and 10 times better than the {\it ROSAT} PSPC.
The SIS was operated in 1-CCD, 2-CCD or 4-CCD
modes, depending on the observation (see Table 3). Only chip n.1 of SIS0
and chip n.3 of SIS1 were used in the analysis of the SIS data, except for
the AO1 observation of S5 0014+81 where also chip n.2 of SIS1 was used because the source
photons were spread equally over the two detectors.
Following a software-related problem on board {\it ASCA}, the data collected
from the GIS3 during the observation of PKS 0537-286 were damaged. They could not be
recovered and therefore were excluded from the analysis.
All observations were performed in FAINT mode and were 
corrected for dark frame error (DFE) and echo uncertainties (Otani \& Dotani, 1994).
The data were selected according to standard (rather
conservative) criteria, i.e., when the angle between the target and the Earth's
limb was $>$ 5$^{\circ}$ (SIS and GIS), when the geo-magnetic
rigidity was $>$ 7 GeV/c (GIS) and $>$ 8 GeV/c (SIS) and when the angle between
the target and the day-night terminator was $>$ 25$^{\circ}$ (SIS). Rise-time
rejection was applied to the GIS data, and hot and flickering pixels were
removed from the SIS data. Telemetry drop-outs and spikes were
excluded from the light curves for each instrument.

Source counts were extracted from circles centered on the sources of
6$^{\prime}$ for the GIS and 3$^{\prime}$ for the SIS. For the SIS instruments,
the background spectra were obtained from the edges of the same CCD chip. The use
of SIS blank sky files for the background yielded spectral results within the
errors reported in the following analysis.
For the GIS instruments, we found that the two standard background 
substraction methods (backgrounds extracted from the blank sky files 
or locally) gave systematically different results for the weakest sources.
Therefore, a non-standard background extracted from the blank sky files was
adopted. Detailed explanation and justification for this choice is given
in the Appendix A.

The relevant data for the {\it ASCA} observations are given in Table 3.
Data preparation and spectral analysis were performed using version 1.0h
of the XSELECT package and version 8.50 of the XSPEC program (Arnaud et al. 1991).

\section{Results}

\subsection{{\it ROSAT} Temporal and Spectral Analysis}

The data were first binned in 400 s time intervals, as suggested by the
wobble period of the telescope, and light curves were plotted 
to evidence variability. However, no significant variation
was detected. Source spectra were thus accumulated for each observation,
And binned with a signal to noise ratio from 4 to 13, depending on
the source statistics.
Three quasars (PKS 0438-436, S5 0836+71 and
PKS 2126-158) had multiple observations. Comparisons between different
observations indicate a clear flux variation for S5 0836+71, and possibly
for the other two quasars. These are discussed in detail in \S 4 where
results for individual sources are presented.
After preliminary analysis, no significant spectral variability was
detected between the multiple observations. These were, therefore,
fitted simultaneously, tying together the fit parameters, but with
the normalizations free to vary.

Fits were performed using a single absorbed power law model with absorption
cross-sections and abundances from Morrison \& McCammon (1983).
The resulting spectral photon
indices $\Gamma$, column densities $N_{\rm H}$ and normalizations
are given in Table 4.
The two-dimensional $\chi^2$ contour plots in the parameter space $N_{\rm H}$--$\Gamma$
are shown in Fig. 1, together with the {\it ASCA} contours (see \S 3.2).
The present re-analysis yielded results consistent with previous measurements (see
references in Table 2).
It is worth pointing  out that the absorption in excess of the Galactic value previously
reported for PKS 0438-436 and PKS 2126-158 (Wilkes et al. 1992, E94) is
confirmed by the present analysis (see Fig. 1 or Table 4), which makes use of two
new PSPC observations of PKS 2126-158 almost doubling the total counts from that
source. The weighted mean photon index for the PSPC sample is
$< \Gamma_{0.1-2.4 \rm keV} >$ $\simeq$ 1.53 $\pm$ 0.06.

More complex models were not attempted since a) the main purpose
of the present analysis of {\it ROSAT} observations is to compare and,
whenever possible, combine these data with the {\it ASCA} data and b) in all cases, a
single absorbed power law model provides an acceptable description
of the spectra. More complex models might, however, be found in the references
given in Table 2.

\subsection{{\it ASCA} Temporal and Spectral Analysis}

Source plus background light-curves were accumulated for each source, and none
of these indicated significant variability. This is not surprising given the large
statistical scatter of the data due to the low counting rates.
GIS and SIS spectra were binned with more than 20 counts/bin
between $\sim$ 0.7--10 keV and $\sim$ 0.5--10 keV, respectively.
The matrices used were the gis[23]v4-0.rmf released in June 1995 for the GIS and
the ``rsp1.1alphaPI'' matrices released in October 1994 for the SIS.
Since the spectral parameters obtained by separately fitting the four detectors
were all consistent at a $\sim$ 90\% confidence level, the data
were fitted simultaneously from all four instruments with the same
model, tying the fit parameters together but allowing the relative normalizations
of the four datasets to vary. 

The spectra were first fitted using a single absorbed power law model with all
parameters free to vary. The resulting best-fit parameters are given in Table 5, together
with the absorbed 2--10 keV flux and intrinsic 2--10 keV luminosity derived from
the fits. In each case, a single absorbed power law model provides an acceptable description
of the spectra. The two-dimensional $\chi^2$ contour plots in the parameter
space $N_{\rm H}$--$\Gamma$ are shown in Fig. 1, together with the {\it ROSAT} contours.
Contours representing the 68\%, 90\% and 99\% confidence limits for
two interesting parameters are indicated for the simultaneous fit
of the GIS and SIS data. The 90\% confidence contours obtained from separately fitting
GIS and SIS are indicated as well. They clearly show that
the spectral parameters obtained from the GIS always agree with the SIS
results at least at a 90\% confidence level.
The elongated shape of the GIS contours in the direction of low column densities
is a consequence of the reduction  of the GIS effective area at low (E $\lsimeq$ 0.8 keV)
energies, where the effect of the absorption is highest; 
to be compared with the SIS detectors sensitive down to E $\sim$ 0.4--0.5 keV.
The much smaller phase space occupied by the {\it ASCA} $\chi^2$ contours compared with
{\it ROSAT}, in particular for S5 0014+81 and NRAO 140, indicate the higher capabilities
of {\it ASCA} to constrain both the photon index (because of the larger energy band) 
and column density (because of the higher statistics).

The contours also clearly show that at least 6 quasars: S5 0014+81, NRAO 140,
PKS 0438-436, S5 0836+71, PKS 2126-158 and PKS 2149-306, have $N_{\rm H}$ values
larger than the Galactic one, at a more than 99\% confidence level.
For the three Parkes sources and NRAO 140, the present results confirm previous 
findings obtained with the same {\it ASCA} data (Serlemitsos et al. (1994), 
S96, Turner et al. 1995).
The most important results are instead for S5 0014+81 and S5 0836+71
for which excess absorption is detected for the first time, at a high
($>$ 99.99\%) confidence level (Fig. 1).
The previous analysis of the {\it ASCA} observation of
S5 0014+81 by Elvis et al. (1994c) didn't show any excess absorption
as only the GIS data had been analyzed, thus
sensitivity at low energies to detect the excess absorption was lost.
A striking result from the present analysis comes from comparing the {\it ROSAT}
and {\it ASCA} contours for S5 0836+71 (Comastri, Cappi \& Matsuoka 1996)
which indicates a large variation in the absorption column density of 
$\sim$ 8 $\times$ 10$^{20}$ cm$^{-2}$ (see \S 4.6 for details). 

It should be noted, however, that although the $\chi^2$ contour plots of Fig. 1
show that the {\it ASCA} and {\it ROSAT} 
parameters are always consistent within their 90\% confidence levels
(except for S5 0836+71), the same contours also indicate
that the best-fit column densities obtained with {\it ASCA}
are systematically larger than the {\it ROSAT} values. 
There are several possibilities to explain this apparent contradiction: 
(1) all quasars experienced a real increase in the absorption column densities 
(between $\sim$ 6--15 $\times$ 10$^{20}$ cm$^{-2}$); (2) all the excess 
absorption is due to SIS calibration uncertainties at low energies; 
(3) the absorption measured by {\it ASCA} (corrected for the calibration 
uncertainties estimated below) is real and the difference between the {\it ASCA} 
and {\it ROSAT} results is due to a) the increased statistic obtained with 
{\it ASCA} or b) the fact that the absorption is more complex than the adopted 
one (i.e. a warm absorber).
Absorption variability (1) can be disregarded as there is no reason to expect 
such increase in all quasars. In order to quantify any systematic instrumental 
effect (2) of the SIS at low energies, a series of tests described
in detail in Appendix B have been applied.
Results from this study indicate that part ($\sim$ 2--3 $\times$ 10$^{20}$ cm$^{-2}$)
of the excess absorption column density measured in
the quasars can indeed be attributed to remaining SIS calibration uncertainties.
However it is very unlikely that the effect is all instrumental since
even after considering a conservative systematic error of $\sim$ 3--4 $\times$
10$^{20}$ cm$^{-2}$ (estimated by us in the Appendix B and, independently,
by Hayashida et al. 1995), the measured column densities are still significantly
higher than the Galactic values. Also in the light of the fairly good agreement between
the {\it ROSAT} and {\it ASCA} results, in particular for PKS 0438-436 and
PKS 2126-158 (see Fig. 1), we are inclined to interpret these excess absorptions
as real.
Either the improved {\it ASCA} signal to noise ratio or a complex absorber, 
therefore, are the most plausible and will be 
further considered below (see \S 3.3.2).

The weighted mean photon index obtained for the whole sample in the observed 
$\sim$ 0.5--10 keV is $<\Gamma_{0.5-10 \rm keV}>\ \simeq$ 1.59 $\pm$ 0.01 with a dispersion $\sigma_{0.5-10 \rm keV}\ \simeq$ 0.10 when
N$_{\rm H}$ is left free to vary and 1.45 $\pm$ 0.01 with a dispersion
of 0.08 when N$_{\rm H}$ is fixed at the Galactic value.
However, fits with the absorption fixed at the Galactic value (Table 5)
are not acceptable for these quasars because clear depressions appear at low
($\lsimeq$ 1 keV) energies in the residuals of SIS and, in some case, of GIS.
The mean slope and dispersion have also been computed, jointly,
using a maximum-likelihood technique (see Maccacaro et al. 1988) 
which has the advantage, over a simple algebraic mean, of weighting 
the individual photon indices according to their measured errors
assuming a Gaussian intrinsic distribution of spectral indices.
The results obtained in this way are $<\Gamma_{0.5-10 \rm keV}> 
\simeq 1.61 \pm 0.04$ and $\sigma_{0.5-10 \rm keV} = 0.10 \pm 0.03$, where the
confidence intervals are the 68\% level for two interesting parameters.
Fitting the data only between 2--10 keV where the effect of the absorption
is negligible gives $<\Gamma_{2-10\rm keV}>$ $\simeq$ 1.53 $\pm$ 0.05 and
$\sigma_{2-10\rm keV}\ \lsimeq \ 0.12$.

The {\it ASCA} spectra were also investigated for the presence of Fe K${\alpha}$
emission lines. Given the high redshifts of the quasars, Fe K${\alpha}$ lines
are expected at energies between $\sim$ 1.5 and 3 keV, where GIS and SIS effective
areas and resolution are highest. No significant Fe K${\alpha}$ emission lines were
detected, with upper limits for the equivalent width of a narrow Gaussian
line at 6.4 keV ranging between 40 and 415 eV (quasar frame, Table 5).
Very similar upper limits were obtained for a line emitted at 6.7 keV
in the quasar frame. It is emphasized, however, that since no constraint can be given on 
the lines widths, the upper limits strongly depend on the assumed widths of 
the lines and become a factor two worse 
than given here if the widths are as large as 0.5 keV.
Next, spectra were searched for neutral Fe K edges at
7.1 keV (quasar frame) and, for ionized  Fe K edges, at the mean energy of 7.8 keV
observed in Seyfert 1 galaxies (Nandra \& Pounds 1994).
Whenever these spectral features were expected in the energy range of 1.8--2.4 keV,
a conservative systematic error of 10 eV (for the lines) and $\Delta \tau$ = 0.05
(for the edges) were added to take into account the uncertainties of the
{\it ASCA} response at these energies (Table 5).
No absorption edges were detected except for S5 0014+81.
For this quasar, the inclusion of an absorption
edge at E $\simeq 1.62$ $\pm$ 0.1 keV (E $\simeq 7.1$ $\pm$ 0.4 keV in the
quasar frame) with depth $\tau\ \simeq \ 0.15^{+0.12}_{-0.11}$ corresponds to
a $\Delta \chi^{2}$ $\simeq$ 9 which is significant at more than 90\% confidence.
The detection is also supported by the fact that the
edge seems to be present in both observations (October 93 and October 94),
at 68\% and 99\% confidence level, respectively.
The equivalent hydrogen column density derived from the edge
is $\sim \ (4-39) \ \times\ 10^{23}$ cm$^{-2}$ (assuming a spherical distribution of the 
absorber and a Fe cosmic abundance
relative to hydrogen of $\sim\ 3 \ \times 10^{-5}$) which is consistent
with $\sim\ (4-7) \ \times\ 10^{23}$ cm$^{-2}$ measured assuming
an intrinsic origin for the absorber (Table 6a).
The predicted Fe K$\alpha$ line equivalent width is in the range of
$\sim$ 40--350 eV (Makishima 1986, Leahy \& Creighton 1993) consistent
with the computed upper limits.

\subsection{{\it ROSAT} and {\it ASCA} Combined Temporal and Spectral Analysis}

For five of the nine quasars: S5 0014+81, NRAO 140, PKS 0438-436, PKS 0537-286 and
PKS 2126-158, the best-fit spectral parameters derived from {\it ROSAT} and
{\it ASCA} data are consistent with each other at a 90\% confidence level.
Despite the fact that the {\it ASCA} spectra require systematically
more absorption than the {\it ROSAT} spectra (see Appendix B), the 0.1--2 keV flux
extrapolated from the {\it ASCA} spectra (Table 5) is in good (within $\sim$ 10\%)
agreement with the measured {\it ROSAT} flux (Table 4). 
Therefore, the data from the two instruments were combined to
benefit of the higher sensitivity of the {\it ROSAT} PSPC at lower energies
and of the {\it ASCA} instrument higher energy resolution and broad band.
The normalizations of the 2 instruments were left free to vary independently.
The unprecedented quality (for high-z quasars) of these combined spectra,
covering typically an energy range of $\sim$ 0.4--40 keV
in the quasar frame, 
provide an excellent opportunity to test the data against more
complex emission models such as complex absorption, thermal emission
and reflection models, as described in the following.

\subsubsection{Single Power-Law Fits}

Again, we first fitted the combined spectra with a single absorbed power
law model, with absorption abundances and cross-sections from Morrison \& Mc
Cammon (1983). The results from these fits are given in Table 6a.
Spectra, residuals and contour plots are shown in Fig. 2. The weighted mean of the photon
indices for the 5 quasars is $< \Gamma_{0.1-10 \rm keV} >\ \simeq$ 1.62 $\pm$ 0.02, 
with a dispersion
$\sigma_{0.1-10 \rm keV}\ \simeq\ 0.11$. Note that, when compared with {\it ASCA} results
alone (Table 5), the addition of {\it ROSAT} data for S5 0014+81 and NRAO 140
gives almost unchanged results. However, for the other
three sources: PKS 0438-436, PKS 0537-286 and PKS 2126-158, there is indeed
a significant improvement combining the data.
Note that the column density for PKS 0537-286 turns out to be marginally
($>$ 90\%) higher than the Galactic value since statistical errors have been
reduced, suggesting excess absorption in this source too. However, considering
the {\it ASCA} SIS systematic uncertainties (Appendix B), conclusions on
this issue are unwarranted.

The inclusion of an Fe K emission line or Fe K absorption edge in the model
gave results almost identical to the one obtained in \S 3.2 and given in
Table 5.

\subsubsection{Absorption Fits}

Given sufficient energy resolution, it should in principle be possible
to constrain the redshift and metal abundances of a neutral absorber through 
direct spectral fitting. 
Though such measurements sound difficult with present generation X-ray
telescopes, it might be worth trying them with the best available data.

As a first step, therefore, elemental abundances were fixed at the cosmic value, given by
Morrison \& Mc Cammon (1983), and spectral fittings with two separate
absorbers, one at z=0 fixed at the Galactic value, and one at the quasar redshift with
$N_{\rm H}$ free were repeated. The resulting parameters are given in Table 6a.
Accordingly, the column densities obtained for an absorber intrinsic to the quasar are
larger, ranging between $\sim$ 1.7--55 $\times$ 10$^{21}$ cm$^{-2}$.
If the redshift of the absorber is left free to vary,
no preferential solution is found 
for any of the quasars indicating that the present data do not allow to distinguish
between a local absorber 
(at z=0) or any other absorber placed between us and the quasars.
Fixing the absorber at the quasars redshifts, if the abundances of the 
metals (C, N, O, Ne, Mg, Si, S, Ar, Ca, and Fe) 
were tied together and left free to vary (increasing by one the number of degrees 
of freedom), no relevant constrain was obtained on the abundances. 
This indicates that the model 
was too complex for the present data and no information on the metal abundances 
of the absorber could be obtained.

The spectra were then fitted again with a single absorbed power law model but this
time the Morrison and McCammon cosmic abundances for the photo-electric
absorption (placed at z=0) were replaced with
more recent abundances values obtained by Anders \& Grevesse (1989) and Feldman (1992).
The two cases gave systematically lower absorption column densities by up to
$\sim$ 12\% and 10\%, respectively. Since these
differences were, in every case, within the reported statistical errors,
they do not substantially modify the present conclusions. This systematic
difference should, however, be kept in mind in particular for future,
more sensitive, X-ray missions.

It should be pointed out, however, that the absorber could be 
ionized (Krolik \& Kallman 1984) as is commonly observed in 
Seyfert 1 galaxies (e.g. Otani 1995).
If present, a warm absorber should imprint two main characteristic features 
in the low energy spectrum: absorption edges at  $\sim$ 0.7--1 keV (quasar frame)
due to OVII and OVIII and extra emission below $\sim$ 0.7 keV
due to the reduced warm absorber opacity at low energies. None of these
features are observed in the present data. However, this is not
surprising since, for a quasar at z = 2, the energy range for the edge
features is redshifted to $\sim$ 0.2--0.4 keV which falls outside the
{\it ASCA} energy range and the extra emission is expected below $\sim$ 0.2 keV
where the PSPC response drops down rapidly.
Moreover, since the present spectra indicate that the {\it ROSAT} spectra were 
typically less absorbed than the {\it ASCA} spectra (\S 3.2), this may be 
somehow the signature of a warm absorber. In order to test this hypothesis, 
a warm absorber model (``absori'' in the XSPEC package) placed at 
the quasars' redshift has, therefore, been directly fitted to the data. 
Free parameters were the ionization parameter ($\xi$ = L/nR$^{2}$ (ergs s$^{-1}$ cm), 
where n is the warm gas density and R the distance of the absorber from the source), 
and the warm column density $N_{\rm W}$ (cm$^{-2}$).
A temperature of 3 $\times$ 10$^{4}$ K (Reynolds \& Fabian 1995) 
and Fe cosmic abundances were assumed.
Results are reported in Table 6b. Though the fits are generally statistically worse 
than with a single absorbed power law model, the data accumulated to date 
do not rule out this model and are consistent with ionization parameters of the 
order of $\sim$ few tens up to few hundreds ergs s$^{-1}$ cm with warm column densities 
ranging between $\sim$ 2--7 $\times$ 10$^{22}$ cm$^{-2}$.

\subsubsection{Alternative continuum models}


The possibility that other continuum models could be applied to the data was
considered in the attempt to explain the low energy cut-off with Galactic
absorption only and/or provide an alternative description of the higher energy
power law spectrum.

A model consisting of a broken power law with the absorption column density
fixed at the Galactic value was first applied in order to check whether a
flat power law is able to describe the observed low energy cut-off.
The spectral curvature may, indeed, be a property of the intrinsic emission
of the quasars, as a break in the continuum is expected if the emission 
is due to the synchrotron mechanism where radiative losses are likely
to steepen the spectra at higher energies. Similar arguments are often used
to explain the convex shape of the X-ray spectra of BL Lac objects 
(e.g. Barr, Giommi \& Maccagni 1988).
The results of these fits are shown in Table 6b. In order to obtain acceptable
fits, the soft component is required to be very flat with typical values of
the photon index $<$ 1, while the photon index of the
hard component is basically identical to the values found with a single
power law model. In the four cases where the break energies are constrained, 
they are found to be in the $\sim$ 0.7--1.5 keV energy range
in the observer frame. 
These fits are statistically acceptable and, in principle, very flat spectra 
could be explained, for example, in the framework of inverse Compton 
models assuming a particular energy distribution of the electrons population 
(i.e. with a convex shape produced by the effect of radiative cooling at high energy 
and escape at low energies (Ghisellini, private communication)). This model therefore 
cannot be directly ruled out.
However, considering that the broken power law model requires an extra free-parameter
and that, statistically, it is never significantly better than the single absorbed 
power law model, using this model is unjustified with the present data.

The data were then fitted with a thermal bremsstrahlung model.
With the absorption fixed at the Galactic value, this model is also not
acceptable because there are significant and systematic
deviations in the residuals at low (E $\lsimeq$ 1 keV, observer frame) energies.
With free absorption, we find that the spectra can also be described 
by a very high temperature (kT $\sim$ 20--90 keV in the source frame) thermal model
and an absorption slightly reduced (but still significantly ($>$99\%) higher than the 
Galactic column) when compared to the single absorbed power 
law model (Table 6c). The fits are, however, in all cases worse than with a single power law
model. Note, moreover, that at these temperatures, a bremsstrahlung 
emission model is virtually indistinguishable from a flat power law 
in the {\it ASCA} energy range and as already noted by S96, these temperatures 
correspond to $\sim$ 10--15 keV in the observer frame, which is suspiciously 
similar to the {\it ASCA} higher energy limit. Since the temperature
of bremsstrahlung emission is primarily determined by its high energy cut-off,
only relatively poor constraints can be set with {\it ASCA} on this parameter (errors 
of $\sim$ 25--90\%, see Table 6c). As a result, the description given with the power law
model is preferable, and the thermal model will not be
considered any further.

Previous {\it Ginga} results (Nandra \& Pounds 1994) have shown that the canonical
X-ray slope ($\Gamma  \sim$ 1.7) of Seyfert 1 galaxies can be interpreted as
the sum of an intrinsic steep power law with $\Gamma_{\rm 2-20 keV}\ \sim\ 1.8-2.1$
plus a reflection component (e.g. Lightman \& White 1988).
Prompted by these results and the fact that
the present sample of quasars exhibit fairly flat ($\Gamma\ \sim\ 1.5-1.7$) spectra
up to $\sim$ 30--40 keV (rest frame), we searched for evidence of high-energy flattening.
A reflection component was included into the absorbed power law model leaving the
absorption and photon index free to vary.
In its simplest form, this model (plrefl in XSPEC) adds only one free parameter:
the relative normalization, R (=A$_{\rm refl}$/A$_{\rm pl}$), between the
reflected component (A$_{\rm refl}$) and the incident power law (A$_{\rm pl}$) (see 
Cappi et al. 1996 for more details on this model).
The reflected component was integrated over all viewing angles.
Present results give no evidence for a reflection component, with upper limits of R ranging
from $\sim$ 0.3 to $\sim$ 0.7, at 90\% confidence limit for two interesting
parameters (Table 6c). The lack of Fe K emission lines further supports this conclusion 
though it should be reminded that broad lines cannot be excluded from the data.


In order to test the consistency of the present results with what is commonly found
in Seyfert 1 galaxies, we forced the intrinsic power law to be steep, with 
a photon index fixed at $\Gamma$ = 1.9 and fitted the data again.
As a result, the model spectra required a substantial amount of reflection,
R $\sim$ 1.1 to 3.1, in order to explain the observed flat spectra.
However, since all the fits became significantly worse ($\Delta\chi^2\ \sim$ 10--20),
the presence of such a reflection component can be ruled out on statistical basis.

On the other hand, a high energy cutoff has been discovered with GRO {\it OSSE} in the 
intrinsic spectrum of Seyfert galaxies which can be described by an exponential 
law of the form E$^{-\Gamma}$ exp(-E/E$_{\rm c}$) (Jourdain et al. 1992) but 
where the precise value of the $e$-folding energy, E$_{\rm c}$, is still 
uncertain ($\sim$ 40--300 keV). A hard X-ray cutoff was therefore added 
to the model and fitted to the data but no constrain was obtained for the $e$-folding energy,
with lower limits on E$_{\rm c}$ ranging between 15--40 keV in the quasar frame.


\section{Comments on Individual Objects}

\subsection{S5 0014+81}

S5 0014+81 is the farthest and optically brightest quasar in this sample. It was
observed in X-rays by {\it EXOSAT} in 1984 (Lawson et al. 1992), by the
{\it ROSAT} PSPC in 1991 (E94, Bechtold et al. 1994)
and by {\it ASCA} in 1993 and 1994.
The GIS data obtained from the first of the two {\it ASCA} observations are
discussed by Elvis et al. (1994c).
Their results are in very good agreement with the present GIS results.
The {\it ROSAT} image indicates that
there are two X-ray sources within $\sim 6^{\prime}$ of
the quasar. One source at about 1.8$^{\prime}$ from the quasar was excluded
from the {\it ROSAT} analysis and was estimated to be negligible in the {\it ASCA}
energy band. The other source, a $V$ = 8.8 K0 star at $\sim$ 5$^{\prime}$
from the quasar was neglected for both instruments.
The photon index and column density derived here (Table 5) are consistent
with previous measurements.
Considering the instrumental uncertainties and different instrumental
band-passes, the {\it EXOSAT} and {\it ROSAT} PSPC spectra are consistent in both
spectral index and flux to the {\it ASCA} results.
However, it is the first time that the column
density is clearly constrained at a value higher than the Galactic column
at a high ($>$99.99\%) significance level (Fig. 2).
The absolute amount of excess absorption is $\sim$ 13.4 $\times$ 10$^{20}$ cm$^{-2}$.
It should be noted that S5 0014+81 is well-known for its Lyman Limit
absorbers which have been object of extensive studies (e.g. Steigman 1994 and
references therein), but that
no damped Ly$\alpha$ system is known along the line of sight to this quasar
(Lanzetta, Wolfe \& Turnshek 1995).

\subsection{PKS 0332-403}

This object was observed in the {\it ROSAT} All-Sky Survey (RASS) with a flux between
0.1--2.4 keV of 1.75 $\pm$ 0.6 $\times$ 10$^{-12}$ erg cm$^{-2}$ s$^{-1}$ (Brinkmann,
Siebert \& Boller 1993). This is in reasonable agreement with the
{\it ASCA} flux of $\sim$ 0.96 $\pm$ 0.1 $\times$ 10$^{-12}$ erg cm$^{-2}$ s$^{-1}$
extrapolated between 0.1--2.4 keV and corrected for absorption.
The {\it ASCA} observations were studied by S96, who derived spectral parameters
consistent with our results.

\subsection{NRAO 140}

NRAO 140 has a long history of X-ray observations. Previous {\it HEAO-1}, {\it Einstein},
{\it EXOSAT} and {\it Ginga} observations have shown that its X-ray spectrum is well
described by a single absorbed power law with $\Gamma\ \sim$ 1.5--1.8 (Marscher 1988,
Ohashi et al. 1992). This source is known for showing a large and variable
X-ray absorption column ($N_{\rm H}\ \sim$ 3--20 $\times$ 10$^{21}$ cm$^{-2}$) which
has been interpreted as the passage of Galactic dense clouds across the line of
sight (Bania, Marscher \& Barvainis 1991). {\it ROSAT} and {\it ASCA} observations
are discussed in Turner et al. (1995). Their combined fit yields 
$\Gamma\ \simeq$ 1.73 $\pm$ 0.03
and $N_{\rm H}$ $\sim$ 3 $\pm$ 2 $\times$ 10$^{21}$ cm$^{-2}$ which is nearly identical
to the present results (Table 6A). The X-ray absorption column is significantly
larger than the Galactic HI column density ($N_{\rm HI}\ \simeq$ 14.2 $\times$ 10$^{20}$
cm$^{-2}$; Elvis et al. 1989) derived from measurements at 21 cm. However, $^{12}$CO
emission measurement toward NRAO 140 implies a molecular hydrogen column density of
($N_{\rm H_2}\ \simeq$ 17 $\times$ 10$^{20}$ cm$^{-2}$) which, when added to the HI column
density, is in excellent agreement with the observed X-ray column (Bania et al. 1991; Turner et al. 1995). These results confirm that the excess
and variable absorption in NRAO 140 plausibly originates from Galactic molecular clouds 
passing accross the quasar line of sight.
It should be noted, however, that these results are based on the assumption that the
CO-to-H$_2$ conversion factor is $\sim$ 3 $\times$ 10$^{20}$ cm$^{-2}$ K$^{-1}$ km$^{-1}$ s,
as commonly found for clouds in the Galactic plane. But this value is rather uncertain
in the case of high-latitude molecular clouds such as those toward NRAO 140.
Assuming the much lower conversion factor of $\sim$ 0.5 $\times$
10$^{20}$ cm$^{-2}$ K$^{-1}$ km$^{-1}$ s obtained by de Vries, Heithausen \&
Thaddeus (1987) or
Heithausen et al. (1993), the molecular hydrogen column density is reduced by a factor
of $\sim$ 6 and implies, again, excess absorption. It should also be noted 
that absorption intrinsic to the quasar is not ruled out by the current data.

\subsection{PKS 0438-436}

The {\it ROSAT} observations of this object were discussed in great detail by
Wilkes et al. (1992) and E94 and its spectral energy distribution
is shown in Bechtold et al. (1994). As E94 noted, the apparent flux decrease by
$\sim$ 30\% between the two {\it ROSAT} observations (Table 4) can be explained
if one takes into account the different wobble mode between the observations. Preliminary
results of the {\it ASCA} observation are given in Serlemitsos et al. (1994) and are
consistent with those presented here.

\subsection{PKS 0537-286}

The {\it ROSAT} and {\it ASCA} observations of PKS 0537-286 were previously discussed by
B\"uhler et al. (1995) and S96, yielding results consistent
with the present analysis. As shown in \S3.3.1, combining the {\it ROSAT} and {\it ASCA}
spectra, there is marginal evidence for excess absorption also in this quasar, 
though, within  the uncertainties in the low-energy response of 
the SIS. Note, however, that even after removing all the SIS data below 1.5 keV, 
the excess absorption remains statistically significant at $>$ 90\% confidence level.

\subsection{S5 0836+71}

As Brunner et al. (1994) noted, S5 0836+71 underwent a flux decrease by a factor of
$\sim$ 2 between the two {\it ROSAT} observations (Table 4), with no evidence of
spectral variation. The time scale of the variation corresponds to $\sim$ 0.2 yr in
the quasar frame. The present analysis shows that the {\it ASCA} flux is consistent
with the {\it ROSAT} flux in the lower state. Compared with the {\it ROSAT} spectra,
the {\it ASCA} results provide strong evidence for variable absorption
($\Delta N_{\rm H}\ \sim 8$ $\times$ 10$^{20}$ cm$^{-2}$) in the direction of
S5 0836+71 on a time-scale of less than 2.6 yr.
Including in the model the Galactic absorption and assuming that the extra absorption 
is intrinsic
to the quasar, the {\it ASCA} fit gives an intrinsic $N_{\rm H}\ 
\simeq\ 1.18^{+0.42}_{-0.37}$
$\times$ 10$^{22}$ cm$^{-2}$. The change in absorption corresponds then to a 
$\Delta N_{\rm H} \sim$ 1 $\times$ 10$^{22}$ cm$^{-2}$ on a time-scale of $\lsimeq$ 0.8 yr
in the quasar frame.
It should be emphasized that fitting the {\it ASCA} spectra with a broken
power law model absorbed by the Galactic column gives $\Gamma_{\rm soft}$ $<$
0.65, E$_{\rm break}$ 0.96$^{+0.09}_{-0.10}$ keV and $\Gamma_{\rm hard}$
$\simeq$ 1.37$^{+0.04}_{-0.03}$, with a fit formally acceptable
($\chi^2_{\rm red}$/dof $\simeq$ 0.87/502).
However, as in the other cases where this model was applied (\S 3.3.3), this model
is unlikely (but not ruled out) since it requires the soft component to be extremely flat.
The {\it ROSAT} and {\it ASCA} spectra were also fitted simultaneously 
with a single warm absorber model (\S 3.3.2). The fits were, however, 
not acceptable because of large residuals in the {\it ASCA} data between 
0.5--1.5 keV. Even a complex model could not, therefore, explain the spectra 
from the two instruments contemporary. Variable absorption (neutral or 
ionized) is therefore required by the present data.
It is interesting to note that S5 0836+71 has also been observed several 
times with EGRET in the high energy $\gamma$-ray band (Nolan et al. 1996).
The reported spectrum is very steep ($\Gamma \simeq 2.5 \pm 0.5$) regardless of the
(variable) $\gamma$-ray intensity.

\subsection{PKS 1614+051}

PKS 1614+051 was marginally detected with the {\it Einstein} IPC (Wilkes et al. 1994)
and in the RASS (S96). The source is very faint also in the
{\it ASCA} observation with a 2--10 keV flux of $\sim$ 2 $\times$ 10$^{-13}$ erg 
cm$^{-2}$ s$^{-1}$.

\subsection{PKS 2126-158}

PKS 2126-158 was first detected in X-rays with the {\it Einstein} IPC (Zamorani et al. 1981). The {\it ROSAT} spectrum is discussed in E94 and the spectral energy
distribution is given in Bechtold et al. (1994). Note that two additional {\it ROSAT}
observations have been analyzed in the present work which almost doubled the total
number of {\it ROSAT} counts. The {\it ASCA} observations were discussed by Serlemitsos
et al. (1994), who tentatively constrained the redshift of the absorber at z$<$0.4.
However, we cannot reproduce such contours, neither with nor without the addition of
the {\it ROSAT} data. No redshift is preferred from our analysis. This
discrepancy might be attributed to the fact that Serlemitsos et al. use older 
response matrices obtained from the preliminary calibrations of the GIS and SIS instruments.

\subsection{PKS 2149-306}

RASS and {\it ASCA} observations of PKS 2149-307 were discussed in S96.
The comparison of the spectra obtained from the two instruments indicates spectral
variability in this source, most likely interpreted in terms of variable absorption
(see Fig. 2 of S96). The column density found in the present analysis is consistent
with, but slightly higher than, the value obtained by S96.
However, it should be noted that, unlike the S5 0836+71 case,
the extra-absorption is not very large in this source ($\Delta N_{\rm H}\ \sim 4\ \times
\ 10^{20}$ cm$^{-2}$ in S96, $\Delta N_{\rm H}\ \sim 6 \times \ 10^{20}$ cm$^{-2}$ in the
present analysis) and could, therefore, be affected significantly by the SIS calibration uncertainties
discussed in \S 3.2 and Appendix B.

\section{Discussion}

\subsection{Excess Absorption}

The most striking result obtained from the present analysis is that the X-ray spectra of
6 high-z RLQs examined here are absorbed by column densities significantly 
higher than the Galactic value. 
Moreover, since these measurements correspond to the
higher signal-to-noise data, they suggest that absorption is common (maybe ubiquitous)
in high-z RLQs. In this section, we discuss the possible origin for such absorption.

\subsubsection{Line-of-Sight Absorption}

{\it a) Galactic}

Assuming that the extra absorption originates somewhere between the observer
and the quasars, the first thing to examine is whether all of it could be
produced in our Galaxy.
This evaluation is commonly done through radio measurements at
21 cm which have been found to reasonably well trace the total amount
of gas in the Galaxy (Dickey \&  Lockman 1990).
However, the column densities are derived from measurements averaged over fairly
large areas (several tens of arcmin$^2$). Moreover, radio observations only detect
interstellar atomic hydrogen and neither molecular gas (e.g. H$_2$, CO, OH, etc.)
nor dust. So it may happen that, for a given path, the total effective absorbing
column is significantly higher than
that indicated by 21 cm measurements. If one also considers the proper motion 
of molecular clouds, it could well be possible to explain strong and variable absorption
by means of special conditions in our Galaxy alone.
A remarkable example is the case of NRAO 140 which is located at the edge of a molecular cloud
(the IC 348 cloud in the Perseus-Taurus region, Ungerechts \& Thaddeus 1987). As
discussed above (\S 4.3),
when the molecular content along the line of sight is properly taken into account through
$^{12}$CO measurements to calculate the total effective absorption column, it is
possible to account for the large X-ray absorption column measured for this quasar.
Under this hypothesis, the absorption variability may plausibly be due to the passage
of absorbing clouds across the line-of-sight. Given this possibility, we checked
for Galactic CO emission in the direction of all quasars of the present sample (Table 7).
Surveys of local CO emission toward extra-galactic sources did not
detect CO emission in the direction of S5 0836+71 and PKS 2149-306 (Liszt \& Wilson
1993, Liszt 1994). A deep (rms $\sim$ 0.04 K) observation in the direction of
PKS 2126-158 puts another limit on any contribution from Galactic absorption in this
source (Hartmann, private communication).
From Table 7, note that S5 0014+81 is the source with the
lowest Galactic latitude after NRAO 140 which increases the probability for a
contamination from Galactic molecular clouds.
Furthermore, S5 0014+81 is located at the edge of a large molecular cloud complex
(the Polaris Flare)
and CO emission is detected from positions nearby the quasar (e.g. from a point of local
maximum at l=120.50 and b=18.63) over an area of more than 1 deg$^2$
(Heithausen et al. 1993). It is difficult to precisely quantify the contribution of
the cloud to the total column density along the line of sight to S5 0014+81.
However, we believe it is probably negligible in this case because from the CO
emission map shown
in Fig. 5 of Heithausen et al. (1993), the CO line intensity should be $\lsimeq$
0.4 K km s$^{-1}$ in the direction of S5 0014+81.
A rather conservative assumption for the CO-to-H$_2$ conversion
factor of $\simeq$ 2--3 $\times$ 10$^{20}$ molecules cm$^{-2}$K$^{-1}$km$^{-1}$s
(Strong et al. 1988) would yield, therefore, a column density $N_{{\rm H}_2}$ $\simeq$ 1.6
$\times$ 10$^{20}$ atoms cm$^{-2}$, which is in any case negligible if compared
to the atomic hydrogen column density of N$_{\rm H}$ $\simeq$ 13.9 $\times$ 10$^{20}$
atoms cm$^{-2}$ derived from 21 cm observations (Dickey \& Lockman 1990).

Another indirect tracer of the total column of gas along a given path is the emission
in the infrared band, in particular at 100 $\mu$m. The {\it IRAS} survey has shown
that the Galactic IR emission is composed of a diffuse background emission plus several
large, filamentary ``cirrus'' features (Low et al. 1984) predominantly associated
to dust in clouds with column densities of a few $\times$ 10$^{20}$ cm$^{-2}$,
and located even at high Galactic latitudes. Using the {\it IRAS} Faint Source
Survey catalog provided on-line (Moshir 1989) the IR emission within a radius of
6$^{\prime}$ of each quasar was investigated (Table 7). There is no evidence for
contamination from cirrus clouds in these quasars, except
for PKS 2126-158 for which the {\it IRAS} Faint Source Survey maps indicate a probable
contamination (Wheelock et al 1994).
From the {\it IRAS} maps, the 100 $\mu$m flux in the direction of PKS 2126-158 is
estimated to be $\lsimeq$ 1.9 Jy (the maximum value in that region)
which corresponds to a brightness $\lsimeq$ 1.5 MJy sr$^{-1}$ for a typical {\it IRAS}
source covering 3$^{\prime} \times 5^{\prime}$. Adopting a conservative range of
dS$_{100}$/d$N_{\rm H}$ $\sim$ 0.5--2.0 MJy sr$^{-1}$/10$^{20}$ cm$^{-2}$
(Reach , Heiles, \& Koo 1993, Heiles, Reach, \& Koo 1988, de Vries et al. 1987) as typical dust-to-gas
ratios, this implies a total (H$_{2}$ + HI) hydrogen column density of 0.7--3
$\times$ 10$^{20}$ cm$^{-2}$. These values are close to the 21 cm estimate of $N_{\rm H}
\simeq 4.85 \times 10^{20}$ cm$^{-2}$ (Elvis et al. 1989), thus ensure
that there is no significant excess IR emission and therefore that cirrus contamination is
likely to be negligible for the present discussion even in this case.

In summary, on the basis of the above local CO properties and IR emission in the
direction of these quasars, it is unlikely that all the extra absorption measured in
X-rays is attributable to absorption by molecular clouds in our Galaxy, except possibly
for NRAO 140.
However, some contamination from molecular clouds and cirrus clouds
may be present in S5 0014+81 and PKS 2126-158 respectively, but at a low level.

{\it b) Extra-galactic}

It has been known for decades that the space density of quasars decreases above z$\sim$3
(see Shaver et al. 1995 for a review).
Though one plausible explanation for this decrease
may be related to the way quasars evolve, it has also been argued that dust and gas
in intervening galaxies could explain or at least contribute to the apparent turnover
in space density (Ostriker \& Heisler 1984).
Examples of probable intervening galaxies have already been found in several high-z
BL Lac objects (PKS 1413+135: Stocke et al. 1992; AO 0235+164: Madejski 1994; PKS 0537-44
and W1 0846+561: Narayan \& Schneider 1990; MS 0205.7+3509: Stocke, Wurtz \& Perlman 1995).
Interestingly, soft X-ray absorption is often present at a level of $\sim$ few $\times$
10$^{21}$ cm$^{-2}$, which is similar to our findings.


A related possibility is absorption by gas and/or dust in damped Ly$\alpha$ systems
(e.g. Fall \& Pei 1993, 1995) which are plausibly associated to galaxy progenitors
along the line of sight (Wolfe 1995). As extensively discussed by Elvis and
collaborators in a variety of papers (E94 and references therein)
which included three of the present quasars (PKS 0438-436, PKS 2126-158 and S5 0014+81),
the numbers and column densities of intervening damped Ly$\alpha$ systems
(Lanzetta et al. 1995) could possibly explain the measured X-ray absorption.
It should be noted, however, that if the hypothesis of the intervening absorber is correct,
whether it is a galaxy or a damped Ly$\alpha$ system, it predicts that the absorption
should remain constant over long time-scales. This is in apparent contrast with our
findings where significant absorption variability has been detected in S5 0836+71
(\S 4.6) though this ``blazar-type'' object may be intrinsically different 
from the other quasars. 
Moreover, an absorption edge has
been marginally detected in S5 0014+81 at the energy expected if the absorber
is at the quasar redshift. These results suggest that the absorption is more likely
intrinsic to the quasars, as discussed in the following.

\subsubsection{Intrinsic Absorption}

{\it a) Absorption by dust and/or neutral gas}

Recent observations are reviving the long-standing idea (Rieke, Lebofsky \& Kinman 1979,
Sanders et al. 1989)
that quasars themselves may be embedded in large quantities of dust or thick gas.
Webster et al. (1995)
have recently claimed that a large population of radio-loud quasars are so red
that they may have been missed by optical searches. This reddening is interpreted
as arising from dust which is most likely located within the quasar host galaxy
since the effect appears to be independent of redshift.
Other evidence for obscuration have been found in several type 2 AGNs, including
several narrow line radio galaxies of low and high luminosities (e.g. Cen A: Bailey
et al. 1986; 3C 109: Goodrich \& Cohen 1992, Allen \& Fabian 1992; Cygnus A: Ueno et al. 1994; 3C 265: Dey and Spinrad 1996).
In these objects, the absorption is thought to arise in a nuclear torus, with a wide
range of column densities (up to $N_{\rm H}$ $\sim$ 10$^{24-25}$ cm$^{-2}$).
These arguments argue in favor of the presence of large quantities of 
obscuring material, possibly in the form of dust.
Other indirect evidence for large quantities of matter in high-z objects comes from the
recent measurement by Ohta et al. (1996) of a strong CO (5-4) emission line from a
z=4.69 radio-loud quasar. Assuming a Galactic CO-to-H$_2$ ratio, they infer a total
mass of cold gas $\sim$ 10$^{11}$ M$_{\odot}$.

Another possibility for the site of neutral absorbing material is in a cluster of galaxies.
Since RLQs at z$>$0.5 are sometimes located in rich clusters of galaxies
(Yee \& Ellingson 1993), cluster cooling flows may provide the necessary cold material
to produce the observed X-ray absorption (White et al 1991, E94).
The radio galaxy Cygnus-A  is a remarkable example of such scenario
(Reynolds \& Fabian 1996). 
However, again, if the variable absorption observed in S5 0836+71, and possibly PKS 2149-306
is real, this could not be explained either by an absorption torus nor by a cooling flow 
model.
Thus an alternative explanation must be found, at least for these cases.
One possible explanation could be the passage of neutral material (e.g.
dense absorbing clouds) across the radiation field (possibly anisotropic). 
On one hand, cold clouds or ``blobs'' are expected to survive the intense radiation
field close to the central regions of AGNs (Guilbert \& Rees 1988, Celotti, Fabian \&
Rees 1992) and may well imprint reflection features (iron line and hard tail)
commonly seen in the X-ray spectra of AGNs (Nandra \& George 1994). On the other hand,
anisotropic radiation, possibly with beaming of the radiation, may be expected 
in these objects on the basis of the extremely high luminosity (L$_{2-10 \rm keV}$ $\sim$
10$^{47-48}$ erg s$^{-1}$, in the quasar frame) observed and on the basis
of general arguments on the unification of radio-loud quasars (Urry \& Padovani 1995).
As far as S5 0836+71 is concerned, superluminal radio components
(Krichbaum et al. 1990), a rapid $\gamma$-ray flare observed by {\it EGRET} (von Linde
et al. 1993), and theoretical arguments (Dondi \& Ghisellini 1995) support the
hypothesis that, at least in this object, the radiation source is relativistically
beamed. Optically thin clouds with column densities of $N_{\rm H}$ $\sim$ 10$^{21-23}$
cm$^{-2}$ covering the continuum radiation or, alternatively, optically thick
($N_{\rm H}$ $>$ 10$^{25}$ cm$^{-2}$) clouds covering quite a large fraction of it
could plausibly reproduce the observed X-ray absorption. The variability could arise
from the passage of the clouds across the radiation field line of sight.


{\it b) Absorption by ionized gas}

The absorption may also be produced by partially ionized
``warm'' material. Warm absorbers have been previously invoked to explain
the complex X-ray absorption observed in several active galaxies (Nandra \& Pounds
1994; Cappi et al. 1996) and quasars (MR 2251-178: Halpern 1984, Otani 1996; 
3C 351: Fiore et al. 1993, Mathur et al. 1994a; 3C 212: Elvis et al. 1994d, Mathur 1994b).
The state of a warm absorber depends only, though critically,
on the ionization parameter, $\xi$ = L/nR$^{2}$ (Kallman \& McCray 1982).
Since warm absorbers have been unambiguously detected in low luminosity AGNs, they may survive
also at higher luminosities since, for a given density, the gas may just
be at a larger distance from the central source. 
As shown in \S 3.3.2, the present data are consistent with a warm absorber model as 
a possible explanation of the low energy absorption. However, the quality of the data 
coupled with the high redshifts do not allow us to distinguish between a cold and 
ionized absorber. 
It is interesting to note that if the {\it ASCA} X-ray 
spectrum of MR 2251-178 (z $\simeq$ 0.068; L$_{\rm 2-10 keV}$ $\sim$ 10$^{45}$ erg s$^{-1}$), 
which clearly indicates the presence of a warm aborber 
($\xi \simeq 28$, $N_{\rm warm} \simeq$ 2.6 $\times$ 10$^{21}$ cm$^{-2}$),
is redshifted to z=2 (i.e. the data are cut at E $\gsimeq$ 1.2 keV), then it
could be equally well fitted with a neutral absorption column density of
$\sim$ 6--12 $\times$ 10$^{20}$ cm$^{-2}$ (Otani, private communication), similar
to what we find in our sample.
This implies that the data accumulated to date for high-z quasars can hardly be used 
to distinguish between a cold or ionized absorber. A warm absorber cannot, 
therefore, be ruled out.

If the warm absorber hypothesis is correct, variable absorption could be
interpreted as variation in the ionization state of the absorber
(e.g., MCG-6-30-15: Otani et al. 1996). An ionized absorber also predicts
an absorption edge at E $>$ 7.1 keV which
seems to be at variance with our findings for S5 0014+81 where the fitted energy of
the absorption edge is centered at $\sim$ 7.1 keV. The quality of the data,
however, does not exclude mildly ionized absorption either.

\subsection{Statistical Properties}

Correlations among the following parameters: photon
index, column densities, 2--10 keV luminosity, radio-loudness,
and source redshift have been searched for in the whole {\it ASCA} sample.
The analysis shows no significant correlation which
is unsurprising given the small size of the sample. Though not adequate for 
statistical correlations between the X-ray emission parameters, the sample is large 
enough to be representative of the entire class of ``X-ray selected'' high-z RLQs.
But because the sources were selected among the brightest and farthest 
ones detected by previous satellites, the sample is clearly biased toward the 
most luminous objects in the universe.

Earlier X-ray spectra of RLQs, mostly at low redshift, showed mean photon indices
between $\sim$ 1.4--1.8 both in the soft $\sim$ 0.1--4 keV
({\it Einstein} IPC: Wilkes \& Elvis 1987; {\it ROSAT} PSPC: Brunner et al. 1994)
and hard $\sim$ 2--10 keV ({\it EXOSAT}: Comastri et al. 1992; {\it Ginga}: Lawson
\& Turner 1996) energy bands. In some works, significant dispersion around the mean
has been found (Brunner et al. 1994, Lawson \& Turner 1996). If real, this indicates
that either RLQs have an intrinsic dispersion of slopes or the intrinsic
X-ray spectrum is more complex than a single absorbed power law (e.g. soft-excess
or complex absorption). Previous X-ray spectra of high-z RLQs taken with {\it ROSAT}
(E94, Bechtold et al. 1994b) and {\it ASCA} (Serlemitsos et al. 1994,
S96) are consistent with an average photon index between $\sim$ 1.5--1.8
up to 30 keV in the source frame.

The mean photon index derived for the whole sample over the full {\it ASCA}
energy band is $<\Gamma_{0.5-10 \rm keV}>$ $\simeq$ 1.61 $\pm$ 0.04 with a dispersion 
$\sigma_{0.5-10 \rm keV}$ $\simeq$ 0.10 $\pm$ 0.03. 
This average value is in agreement with most of the 
findings described above for low-z and high-z RLQs derived in both the soft
and hard band, and indicates that there is no X-ray spectral evolution
with redshift or luminosity over a broad $\sim$ 1--40 keV energy band
(source frame). This is further illustrated in Fig. 3 which
shows the X-ray spectral slopes observed with {\it ASCA} between $\sim$ 0.5--10 keV
as a function of redshift and 2--10 keV luminosity (quasar frame) together with 
the photon indices obtained with {\it Ginga} for 18 low-z RLQs as reported by
Lawson \& Turner (1996). It is interesting to note that the dispersion is
significantly different from zero also at high-z, though it is
somewhat smaller than the {\it Ginga} value $\sigma$ $\simeq$ 
0.19$^{+0.06}_{-0.04}$ (Lawson \& Turner 1996).
However, the mean photon index obtained with {\it ASCA} in the 2--10 keV energy band,
where the present measurements are not affected by the absorption,
is $<\Gamma_{2-10 \rm keV}>\ \simeq$ 1.53 $\pm$ 0.05, with a dispersion
$\sigma_{2-10 \rm keV}\  \lsimeq \ 0.12$.
The intrinsic dispersion
is, in this case,  consistent with zero within 1$\sigma$. This result implies that
the above non-zero dispersion is probably a consequence of the heterogeneous
absorption measured in the data and that the intrinsic distribution of photon
indices is well characterized by a single slope.

Excluding S5 0836+71 from the sample because of its blazar-like properties, 
the results are as follow: $<\Gamma_{0.5-10 \rm keV}>$ $\simeq$ 1.64 $\pm$ 0.06 
with $\sigma_{0.5-10 \rm keV}$ $\simeq$ 0.07$^{+0.07}_{-0.03}$ and $<\Gamma_{2-10 
\rm keV}>$ $\simeq$ 1.57 $\pm$ 0.09 with $\sigma_{2-10 \rm keV}\  \lsimeq \ 0.14$. 
These are consistent with the above results.


\subsection{Implications for the XRB and GRB}

It is by now widely accepted that AGNs could supply the bulk of the XRB 
emission above a few keV. Detailed modeling, based on the 
unified scheme and assuming the existence of a large population
of absorbed AGNs, provide a good fit to the observed XRB spectrum 
up to $\sim$ 100 keV (Madau, Ghisellini \& Fabian 1994; Comastri et al. 1995).
If the absorption discovered in some of the objects in our sample
turns out to be a common property of high-z, high-luminosity RQQs
this would give a strong support to the above models.

It should be noted that RLQs constitute only a small fraction of 
the quasar population so that
their contribution to the XRB is of the order of a few percent.
This point is also enforced by our findings that the observed X-ray spectra 
of RLQs have $\Gamma \sim 1.5-1.6$ up to $\sim$ 40 keV in the source frame, 
steeper than the XRB as illustrated in Fig. 3.
But the derived mean slope for the present sample has interesting implications 
for the $\gamma$-ray background.
In fact, it has been recently pointed out (Comastri, Di Girolamo \& Setti 1996) 
that flat spectrum radio quasar (FSRQ) may provide a significant 
fraction ($\sim$ 70--80 \%) of the $\gamma$-ray background above several 
tens of MeV.
This result has been achieved assuming that the X-ray spectra 
of FSRQ in the X-ray band (from $\sim$ 1 keV up to several hundreds of keV) 
are characterized by a slope with $\Gamma = 1.5$.
Our findings provide, therefore, further support for this model suggesting that 
RLQs are likely to provide an important 
contribution to the hard X-ray -- $\gamma$-ray backgrounds.


\section{Conclusions}

The present analysis of {\it ASCA} and {\it ROSAT} observations of high-z RLQs confirms
that excess absorption is common in these objects. It is indeed detected significantly
in the six brightest quasars of our sample. Unfortunately, even
combining the two instruments, it is not possible to constrain the redshift of the
absorber through direct spectral fitting. However, new results for S5 0836+71, and
possibly S5 0014+81, favor the hypothesis that, at least in these two cases, the 
absorption is intrinsic to the quasars. Indeed, a comparison of
{\it ASCA} and {\it ROSAT} spectra indicates the absorption has varied
($\Delta N_{\rm H} \sim 8 \times 10^{20}$ cm$^{-2}$) in S5 0836+71 on a
time-scale of $\sim$ 2.4 yrs (observer frame) and an Fe K absorption edge is marginally
detected in the {\it ASCA} observations of S5 0014+81, at the energy
expected if the absorber is at the quasar's redshift.
If we assume that the absorption is intrinsic for all objects in the sample,
the column densities implied range between $\sim$ (1.7--55) $\times$ 10$^{21}$ cm$^{-2}$.

Since the data do not allow to constrain unambiguously the redshift of the absorption,
several possible origins have been considered.
A careful search for Galactic CO and/or
IR (100 $\mu$m) emission in the direction of the quasars indicates that the
contribution to the total X-ray absorption column from molecular gas and dust
present in our Galaxy is negligible for the majority of the quasars. It is, however,
likely to be significant in the case of NRAO 140. Extra-galactic absorption from
intervening galaxies and/or damped Ly$\alpha$ systems may be relevant for some of
the objects, but is not consistent with the new findings for S5 0836+71 and
S5 0014+81. Intrinsic absorption from an absorption torus or intra-cluster gas is
also unlikely in the light of the absorption variability. It is argued, therefore, that
the preferable explanation is intrinsic absorption by cold gas,
possibly in the form of clouds, near the central source. A warm absorber
would be a plausible candidate as well.

The accumulated counts allowed us to put, for most objects, stringent limits 
on the presence
of an Fe K emission line, with upper limits on the EW ranging between $\sim$ 40--400 eV
 in the quasar frame. There is, however, more uncertainty about the strength of the line 
if the width is left free to vary.
Alternative continuum models have also been fitted to the data.
The upper limits on the intensity of Compton reflection range between R $\sim$ 0.1--0.7.
A broken power law model, tested as an alternative explanation for the low energy cut-off,
would require unreasonably flat ($\Gamma < 1$) slopes at low energies, though it cannot be 
ruled out by the data. It is also argued that
a thermal bremsstrahlung origin for the observed continuum is unlikely.

The average photon index ($<\Gamma>\ \sim$ 1.5--1.6) is consistent with previous
measurements at high and low-z, therefore suggesting that RLQs do not show
spectral changes over about two decades in energy ($\sim$ 0.4--40 keV)
with either redshift or luminosity.
This average slope, which is steeper than the spectrum of the XRB but flatter
than the average slopes of RQQs and low-luminosity AGNs, also suggests that RLQs do
not contribute significantly to the XRB but, instead, are likely to provide
a significant fraction of the $\gamma$-ray background flux.

\ms
\centerline{\bf APPENDICES}

\appendix
\section{GIS Backgrounds}


In the standard way of analyzing {\it ASCA} GIS spectra, there are two
possible choices for the background subtraction: either from
a source-free part of the observation field of view (hereafter ``local background''),
or from the available blank sky observations with the same
region filter as used for the source spectrum (see ``The ABC
Guide to {\it ASCA} Data Reduction''). Both methods should, in principle,
give similar results. However, slight but significant differences in the source
spectral slopes were found in the present analysis and forced us to choose
a non-standard background region.

The problem is that background regions extracted from the source field of view 
almost inevitably fall at a substantial off-axis distance because
sources are usually pointed nearly on-axis.
But as the distance from the X-ray telescope optical axis increases,
reduction of the contribution from the cosmic X-ray background (due to the vignetting) and
softening of the non X-ray background (Kubo et al. 1994) change in a
complex and position dependent manner the spectrum of the background, which
cannot be corrected in the current analysis procedure.
On the other hand, standard background regions extracted from the blank sky
files from the same area as
that used for the source almost always include the regions contaminated by the
Seyfert 2 galaxy NGC 6552 (Ebisawa 1994). 
Because of the very strong Fe K emission line of this source at $\sim$ 6.4 keV
(Fukazawa et al. 1994), the source spectra obtained with blank sky backgrounds tend
to be steeper than those obtained with local background subtractions. 
As a result, in the analysis
of the weakest sources (typically those with intensity $\lsimeq$ 0.1 cts/s),
different backgrounds often yielded significantly different source slopes with
deviations up to $\Delta \Gamma \sim 0.2$.
Therefore, we adopted a non-standard background
extracted from the GIS standard
blank sky files from a region non contaminated by NGC 6552 and at the same or, if not
possible, similar off-axis distance of the source region.
This choice of background has the advantage to avoid the above problems
 of the off-axis position with the local background and of the contamination
by NGC 6552 with the blank sky background. However, it has the disadvantage
that the screening criteria of the blank sky background never match exactly
those of the source observation (similarly to any other choice of blank sky background).
However, we estimate this is a minor problem when standard screening criteria
are adopted because the GIS background reproducibility (in time) is achieved
with systematic errors less than a few percent (Ikebe et al. 1995).

\section{On the {\it ASCA} SIS response uncertainties at low energies}


Although the results obtained with {\it ASCA} and {\it ROSAT} always agree
within their statistical errors (except for S5 0836+71), the present
analysis indicates that the {\it ASCA} spectra
require systematically more absorption than the {\it ROSAT} spectra (Fig. 1). This
suggests that the SIS responses may suffer of a systematic excess absorption,
though {\it ROSAT} PSPC calibration uncertainties (e.g.
non-linear gain variations) may play a role as well.
This confirms previous, more detailed, studies on the SIS calibration
uncertainties at low energies (e.g. Hayashida et al. 1995).
Indeed, it is widely recognized that there are local features around 0.5--0.6 keV
which show up, preferentially in strong sources, in the form of absorption edges around
0.5 keV and/or emission lines at $\sim$ 0.6 keV (e.g. Guainazzi \& Piro 1994, Cappi et al. 1996, Dotani et al. 1996).
The origin of the problem is not clear, though probably related to the presence
of the Oxygen K-edge (produced in the CCD dead layer) at E $\simeq$ 0.54 keV in the SIS response.
Because of this edge, a small offset and/or variations of a few eV in the exact response
energy scale could easily be responsible for such features.
Though no clear feature is evident in the low energy residuals of the quasars spectra
(Fig. 2), it is possible that it could have affected somehow our measurements.
Other uncertainties concern the absolute SIS efficiency at low energies which
is currently being quantified in terms of a systematic error of $\sim$ 3 $\times$
10$^{20}$ cm$^{-2}$ for SIS0 and $\sim$ 2 $\times$ 10$^{20}$ cm$^{-2}$ for SIS1
in the absorption column (Hayashida et al. 1995, Dotani et al. 1996). This problem is probably
related to the correct calibration of the SIS quantum efficiency, i.e. the thickness of
the CCD dead layer.
It is not clear, however, how much the above local features at
E $\sim$ 0.5 keV are responsible for and/or related to such excess absorption but,
in principle, both effects (local features and excess absorption) should be already
taken into account in the systematic error of $\sim$ 3 $\times$ 10$^{20}$ cm$^{-2}$.

In light of the above considerations, we have performed a series of checks in order to
establish the reliability of the measured excess absorption for all {\it ASCA} spectra.


{\it a) Test 1 - Artificial gain shifts:} 

Using the ``gain'' command in XSPEC, we shifted the energy scales of the
SIS0 and SIS1 responses between $\pm$ 20 eV, i.e. within the typical systematic 2$\sigma$
uncertainties in the energy scale (Otani \& Dotani 1994).
The observed changes in the column densities were, in every case, lower than
2 $\times$ 10$^{20}$ cm$^{-2}$, which is smaller than the typical statistical
errors found in the present work.

{\it b) Test 2 - Cut of the data below 0.65 keV:}

When, more drastically, all data below 0.65 keV were ignored, the 
fits yielded even larger best-fit absorption columns for the quasars
NRAO 140, PKS 0438-436, PKS 0537-286 and PKS 2126-158. For S5 0014+81 and
S5 0836+71, the columns were lowered by $\sim$ 1 $\times$ 10$^{20}$ cm$^{-2}$
($\sim$ 5\%) and $\sim$ 2.5 $\times$ 10$^{20}$ cm$^{-2}$ ($\sim$ 20\%),
respectively. In every case, the absorption columns remained significantly higher
than the Galactic value at more than 99\% confidence level.

From a) and b), we conclude that the measured SIS excess absorption cannot be attributed
to the local detector features commonly reported around 0.5--0.6 keV. The problem is 
more likely related to the SIS efficiency as a whole at low energies.

{\it c) Test 3 - Comparison with the calibration source 3C 273:}

The SIS response matrices were calibrated on the ground and then cross-calibrated
with the GIS in-flight using the observations of the quasar 3C 273. In particular,
the response matrices of SIS0 chip number 1 and SIS1 chip number 3 (the chips
used in the present analysis) were calibrated using an observation performed
in 1993, December 20$^{\rm th}$. The calibrations were performed taking the
slopes from the GIS spectra and assuming Galactic absorption. 3C 273 was
observed 8 more times by {\it ASCA}. Results obtained from the overall observations
(Cappi \& Matsuoka 1996) do not modify the conclusions presented below. 
For the purpose of investigating the efficiency of the SIS, only the results
from the December 20$^{\rm th}$ 1993 calibration observation are reported below.
The observation was analyzed using the same criteria
(same screening criteria, extraction regions, etc.)
used for the quasars to allow a direct comparison.
About 82 000 counts/SIS were collected in the 0.4--10 keV energy range,
for a total exposure time of $\sim$ 15 500 s.
The spectra extracted for SIS0 and SIS1 were then fitted with a single absorbed
power law, first separately and then simultaneously. Confidence contours
obtained from each fit are shown in Fig. 4.
The best-fit column density is larger than the Galactic column in both SIS0 and SIS1,
yielding an excess column density of $\sim$ 2 $\times$ 10$^{20}$ cm$^{-2}$ in the fit
with both detectors.
This value gives a direct estimate of the systematic error we must take into account
when interpreting the results from the present analysis. This estimate is in 
good agreement with the measurements obtained by Hayashida et al. (1995). Also,
it should be noted that independent analysis of the SIS data extracted from the
central regions of Coma cluster also show excess absorption of $\sim$ 3 $\times$ 
10$^{20}$ cm$^{-2}$ (Hashimotodani, private communication).

Therefore, we conclude that the total systematic error of the absorption column
density should be {\it conservatively} smaller than $\sim$ 3--4 $\times$ 10$^{20}$ cm$^{-2}$.
In the analysis presented in \S 3 and when discussing the results, only deviations
larger than this value have been considered.

\acknowledgments

\pn
{\bf ACKNOWLEDGMENTS}
\ms

The authors would like to thank all the {\it ASCA} team members
for making these observations possible. M.C. thanks T. Dotani, Y. Ikebe, T. Kotani, 
K. Leighly, C. Otani and T. Yaqoob for helpful discussions on the calibration of the SIS
and Dr. Dap Hartmann for measuring the Galactic CO emission in the direction of PKS 2126-158.
M.C. also acknowledges financial support from the Science and
Technology Agency of Japan (STA fellowship), hospitality from the
RIKEN Institute and support from the European Union. G.G.C.P.
acknowledges partial financial support from MURST and ASI.
A.C. acknowledges financial support from the Italian Space Agency under 
the contracts ASI-94-RS-96 and ASI-95-RS-152.
This work has made use of the NASA/IPAC Extragalactic Database (NED)
which is operated by the Jet Propulsion Laboratory, Caltech, under contract
with the National Aeronautics and Space Administration.

\vfill\eject

\onecolumn

\centerline{TABLE 1: The Radio-Loud Quasar Sample}
\ss
\begin{center}
\begin{tabular}{cccccccc}
\hline
\hline
\multicolumn{1}{c}{Object} &
\multicolumn{1}{c}{R.A.$^{a}$} &
\multicolumn{1}{c}{DEC.$^{a}$} &
\multicolumn{1}{c}{z$^{a}$} &
\multicolumn{1}{c}{$N_{\rm Hgal}^{b}$} &
\multicolumn{1}{c}{m$_{\rm V}^{a}$} &
\multicolumn{1}{c}{f$_{5\rm GHz}^{a}$ } &
\multicolumn{1}{c}{R$_{\rm L}^{c}$}\\
\multicolumn{1}{c}{} &
\multicolumn{1}{c}{(J2000)} &
\multicolumn{1}{c}{(J2000)} &
\multicolumn{1}{c}{} &
\multicolumn{1}{c}{($10^{20}$ cm$^{-2}$)} &
\multicolumn{1}{c}{} &
\multicolumn{1}{c}{(mJy)} &
\multicolumn{1}{c}{} \\
\hline
S5 0014+81  & 00 17 08.5 & +81 35 08.1 & 3.38 & 13.9 & 16.5 & 551 & 2.78 \\ 
PKS 0332-403 & 03 34 13.6 & -40 08 25.4 & 1.44 & 1.43 & 18.5 & 2600 & 4.24 \\
NRAO 140 & 03 36 30.1& +32 18 29.3& 1.26& 14.2$^*$ & 17.5 & 2500 & 3.87 \\
PKS 0438-436 & 04 40 17.2 & -43 33 08.6 & 2.85 & 1.47 & 18.8 & 7580 & 4.84 \\
PKS 0537-286 & 05 39 54.3 & -28 39 56.2 & 3.10 & 1.95$^*$ & 20.0 & 990 & 4.39 \\
S5 0836+71  & 08 41 24.4 & +70 53 42.2 & 2.17 & 2.78 & 16.5 & 2573 & 3.45 \\
PKS 1614+051 & 16 16 37.5 & +04 59 33.2 & 3.21 & 4.90 & 19.5 & 850 & 4.15 \\
PKS 2126-158 & 21 29 12.2 & -15 38 40.8 & 3.27 & 4.85$^*$ & 17.0 & 1240 & 3.33 \\
PKS 2149-306 & 21 51 55.5 & -30 27 53.7 & 2.34 & 1.91 & 17.9 & 1150 & 3.66 \\
\hline
\end{tabular}
\end{center}
\pn
$ ^{a}$ Coordinates, redshift, V magnitude and radio flux at 5 GHz,
from V\'eron-Cetty \& V\'eron (1993).
\pn
$ ^{b}$ Galactic absorption from Dickey \& Lockman (1990).
The values marked with (*) are from Elvis et al. (1989).
\pn
$^{c}$ Radio-Loudness defined as R$_{\rm L}$ = log(f$_{5\rm GHz}$/f$_{\rm V}$).
We used m$_V$(0)=3360 Jy (Wamsteker W. 1981)

\bs


\centerline{TABLE 2 : {\it ROSAT} Observation Log}
\begin{center}
\begin{tabular}{ccccccc}
\hline
\multicolumn{1}{c}{Object} &
\multicolumn{1}{c}{Date} &
\multicolumn{1}{c}{Wobble} &
\multicolumn{1}{c}{Matrix} &
\multicolumn{1}{c}{Exposure (s)} &
\multicolumn{1}{c}{NC$^{a}$} &
\multicolumn{1}{c}{Ref.$^{b}$}\\
\multicolumn{1}{c}{}&
\multicolumn{1}{c}{}&
\multicolumn{1}{c}{}&
\multicolumn{1}{c}{}&
\multicolumn{1}{c}{}&
\multicolumn{1}{c}{}&
\multicolumn{1}{c}{}\\
\hline
S5 0014+81&03/15/91&on& DRM06 & 5951&394& 1,2\\
NRAO 140&08/08/92&on& DRM36 & 4039&992& 3 \\
PKS 0438-436& 02/19/91&off& DRM06 &10725&645& 1,4\\
PKS 0438-436&09/19/92&on& DRM36 &10506&547&1\\
PKS 0537-286&09/28/92&on& DRM36 &9487&568& 5 \\
S5 0836+71&03/23/92&on& DRM36 &6993&5400& 6\\
S5 0836+71&11/02/92&on& DRM36 &5026&2008& 6\\
PKS 2126-158&05/08/91&on& DRM06 &3424&613& 1,2 \\
PKS 2126-158&11/12/92&on& DRM36 &3968&729&1,2\\
PKS 2126-158&04/27/93&on& DRM36 &4160&741& This work\\
PKS 2126-158&05/17/93&on& DRM36 &1610&321& This work\\
\hline
\hline
\end{tabular}
\end{center}
\small
\pn
$^a$ NC: Net counts.
\pn
$^b$ References: (1) E94, (2) Bechtold et al. (1994),
(3) Turner et al. (1995), (4) Wilkes et al. (1992), (5) B\"uhler et al. (1995),
(6) Brunner et al. (1994)

\vfill\eject

\centerline{TABLE 3: {\it ASCA} Observations Log}
\begin{center}
\begin{tabular}{cccccccc}
\hline
\hline
\multicolumn{1}{c}{Object}&
\multicolumn{1}{c}{Date} &
\multicolumn{2}{c}{Exposure$^a$ (s)} &
\multicolumn{2}{c}{Count Rate$^a$ ($\times$10$^{-2}$ s$^{-1}$)} &
\multicolumn{1}{c}{CCD mode} &
\multicolumn{1}{c}{Ref.$^b$}\\
\multicolumn{1}{c}{} &
\multicolumn{1}{c}{} &
\multicolumn{1}{c}{GIS} &
\multicolumn{1}{c}{SIS} &
\multicolumn{1}{c}{GIS} &
\multicolumn{1}{c}{SIS} &
\multicolumn{1}{c}{} &
\multicolumn{1}{c}{} \\
\hline
S5 0014+814 (AOI) & 10/29/93 & 39004 & 21750  & 7.33 & 3.88 & 4 CCD& (1) + This work for SIS\\
S5 0014+814 (AOII) & 10/07/94 & 27494 & 20703 & 5.14 & 5.98 & 1 CCD& This work\\
PKS 0332-403 & 08/12/94 & 16646 & 14153 & 1.98 & 2.65  & 1 CCD& (2)\\
NRAO 140 & 02/02/94 & 33181 & 32178 & 16.3 & 20.7 & 1 CCD & (3) \\
PKS 0438-436 & 07/13/93 & 34499 & 26412 & 2.77 & 3.20 & 4 CCD& (4)\\ 
PKS 0537-286 & 03/12/94 & 36236 & 29052 & 2.51 & 4.21 & 1+2 CCD& (2)\\
S5 0836+71  & 03/17/95 & 16532 & 10540 & 24.0 & 30.3 & 1 CCD& This work\\
PKS 1614+051 & 08/02/94 & 39695 & 32577 & 0.88 & 0.55 & 1 CCD& (2)\\
PKS 2126-158 & 05/16/93 & 15914 & 14291 & 10.1 & 12.8 & 1 CCD& (4)\\
PKS 2149-306 & 10/26/94 & 19168 & 16504 & 19.0 & 25.5 & 1 CCD& (2)\\
\hline
\hline
\end{tabular}
\end{center}
\pn
$^a$ Reported values for the GIS and SIS are averaged over the detectors
(GIS2 with GIS3 and SIS0 with SIS1).
\pn
$^b$ References: (1) Elvis et al. (1994c), (2) S96,
(3) Turner et al. (1995), (4) Serlemitsos et al. (1994)

\ss
\centerline{TABLE 4 : {\it ROSAT} X-Ray Spectral Fits}
\centerline{PSPC - Single Power Law ($N_{\rm H}$ free)}
\begin{center}
\begin{tabular}{cccccccc}
\hline
\hline
\multicolumn{1}{c}{Object} &
\multicolumn{1}{c}{Energy Range$^a$} &
\multicolumn{1}{c}{$N_{\rm H}$} &
\multicolumn{1}{c}{$\Gamma$} &
\multicolumn{1}{c}{A$_{\rm pl}^b$} &
\multicolumn{1}{c}{$\chi^{2}_{red}$/dof} &
\multicolumn{1}{c}{F$_{\rm SX}^c$}&
\multicolumn{1}{c}{log L$_{\rm SX}^d$}\\
\multicolumn{1}{c}{} &
\multicolumn{1}{c}{(keV)}&
\multicolumn{1}{c}{($10^{20}$ cm$^{-2}$)} &
\multicolumn{1}{c}{} &
\multicolumn{1}{c}{} &
\multicolumn{1}{c}{} &
\multicolumn{1}{c}{} &
\multicolumn{1}{c}{} \\
\hline
S5 0014+81 &$\sim$0.4-10 &20.0$^{+30.0}_{-15.2}$&1.89$^{+2.40}_{-1.03}$&5.30&1.37/17 & 0.73 & 47.8\\
\\
NRAO 140 &$\sim$0.2-5 &22.5$^{+21.9}_{-14.0}$&1.53$^{+1.14}_{-0.81}$&20.7&1.21/15 & 2.88 & 46.9 \\
\\
PKS 0438-436 (Feb 91)&$\sim$0.4-9 &8.13$^{+20.4}_{-3.69}$&1.71$^{+1.16}_{-0.48}$&3.53&0.45/19 & 0.71 & 47.3\\
\dotfill (Sep 92)&'' &5.07$^{+4.09}_{-2.23}$&1.57$^{+0.49}_{-0.44}$&2.51&1.28/25 & 0.57 & 47.0\\
\dotfill (tot.)  &'' &6.31$^{+4.19}_{-2.04}$&1.63$^{+0.36}_{-0.32}$& &0.93/46 &  & \\
\\
PKS 0537-286 &$\sim$0.4-10 &3.11$^{+1.84}_{-1.50}$&1.50$^{+0.45}_{-0.44}$&2.61&1.84/7 & 0.65 & 47.1\\
\\
S5 0836+71 (Mar 92)&$\sim$0.3-7 &3.29$^{+0.48}_{-0.46}$&1.51$^{+0.13}_{-0.12}$&35.0&1.27/18 & 8.59 & 47.8\\
\dotfill  (Nov 92) &'' &3.45$^{+0.79}_{-0.80}$&1.55$^{+0.22}_{-0.22}$&17.9&1.55/18 & 4.41 & 47.5\\
\dotfill (tot.)    &'' &3.33$^{+0.42}_{-0.41}$&1.52$^{+0.12}_{-0.11}$& &1.34/38 & & \\
\\
PKS 2126-158 (May 91)& $\sim$0.4-10&8.38$^{+13.7}_{-3.82}$&1.44$^{+0.81}_{-0.50}$&10.4&0.95/18 & 2.08 & 47.7\\
\dotfill (Nov 92)&'' &9.01$^{+22.8}_{-4.25}$&1.50$^{+1.22}_{-0.46}$&10.9&0.93/21 & 2.13 & 47.8 \\
\dotfill (Apr 93)&'' &7.23$^{+8.27}_{-2.89}$&1.51$^{+0.57}_{-0.42}$&10.0&0.85/15 & 2.08 & 47.7\\
\dotfill (May 93)& ''&11.5$^{+37.6}_{-6.61}$&1.91$^{+2.07}_{-0.74}$&12.7&1.05/15 & 2.23 & 48.2\\
\dotfill (tot.) &'' &8.42$^{+4.88}_{-2.30}$&1.53$^{+0.34}_{-0.25}$& &0.90/75\\
\hline
\hline 
\end{tabular}
\end{center}
\small
\pn
Note: Intervals are at 90\% confidence, for two interesting parameters.
\pn
$ ^a$ Approximate energy range, in the quasar rest frame.
\pn
$ ^b$ Unabsorbed flux at 1 keV (observed frame) in units of 10$^{-4}$ photons cm$^{-2}$ s$^{-1}$ keV$^{-1}$.
\pn
$ ^c$ Absorbed flux between 0.1--2 keV (observed frame) in units of 10$^{-12}$ ergs cm$^{-2}$
s$^{-1}$.
\pn
$ ^d$ Intrinsic luminosity between 0.1--2 keV (quasar frame) in units of ergs s$^{-1}$.

\vfill\eject

{\centerline{TABLE 5: {\it ASCA} X-Ray Spectral Fits}
\ss
\centerline{GIS+SIS - Single Power Law ($N_{\rm H}$ free; $N_{\rm H} \equiv N_{\rm Hgal}$)}
\begin{center}
\begin{tabular}{@{\hspace{-1.5cm}}ccccccccccc}
\hline
\hline
\multicolumn{1}{c}{Object}&
\multicolumn{1}{c}{Energy Range$^a$} &
\multicolumn{1}{c}{$N_{\rm H}$} &
\multicolumn{1}{c}{$\Gamma$} &
\multicolumn{1}{c}{A$_{\rm pl}^b$} &
\multicolumn{1}{c}{$\chi^{2}_{red}$/dof} &
\multicolumn{1}{c}{F$_{\rm HX}^c$}&
\multicolumn{1}{c}{F$_{{\rm SX}^{\prime}}^d$}&
\multicolumn{1}{c}{log L$_{\rm HX}^e$}&
\multicolumn{1}{c}{EW (FeK)$^f$}&
\multicolumn{1}{c}{$\tau_{\rm edge} (FeK)^g$}\\
\multicolumn{1}{c}{} &
\multicolumn{1}{c}{(keV)} &
\multicolumn{1}{c}{($10^{20}$ cm$^{-2}$)} &
\multicolumn{1}{c}{} &
\multicolumn{1}{c}{} &
\multicolumn{1}{c}{} &
\multicolumn{1}{c}{} &
\multicolumn{1}{c}{} &
\multicolumn{1}{c}{} &
\multicolumn{1}{c}{(eV)}&
\multicolumn{1}{c}{} \\
\hline
S5 0014+81 (Oct 93)&$\sim$2--40 &28.4$^{+6.2}_{-5.7}$&1.75$^{+0.09}_{-0.08}$ &6.46&1.03/314&2.43& 0.82 &47.8&$<$75 & $<$0.23\\
  \dotfill (Oct 94)&''  &26.9$^{+6.9}_{-6.3}$&1.66$^{+0.10}_{-0.12}$& 6.60 & 0.92/237 & 2.49 & 0.73 &47.7 & $<$131 & 0.21$^{+0.18}_{-0.17}$\\
 \dotfill (tot.)& '' &28.1$^{+4.5}_{-4.2}$ & 1.72$^{+0.06}_{-0.07}$ & --- & 1.02/553 & & & &$<$70 &0.15$^{+0.12}_{-0.11}$ \\
                &  &13.9 Fixed & 1.54$^{+0.03}_{-0.03}$& --- & 1.13/554 & &\\
\\
PKS 0332-403 	&$\sim$1--20 &$<$ 22.9 & 1.73$^{+0.28}_{-0.25}$ & 2.48 & 0.77/103 & 0.94 & 0.47&46.3 & $<$415 & $<$0.66\\
		&  &1.43 Fixed & 1.60$^{+0.11}_{-0.12}$ & 2.05 & 0.78/104 & &\\
\\
NRAO 140        &$\sim$1--20 &31.8$^{+3.0}_{-2.8}$& 1.70$^{+0.05}_{-0.04}$ & 21.8 &0.96/864 & 9.09 & 2.56&47.1& $<$36 & $<$0.10 \\
                &  &14.2 Fixed & 1.46$^{+0.02}_{-0.02}$ & 16.3  & 1.23/865 & &\\ 
\\
PKS 0438-436 	&$\sim$2--38 &15.2$^{+9.1}_{-7.7}$ & 1.63$^{+0.16}_{-0.14}$ & 3.89 & 0.93/236 & 1.75 & 0.63 &47.3 &$<$240$^h$& $<$0.29$^h$\\
		&  &1.47 Fixed & 1.42$^{+0.06}_{-0.06}$ & 2.92& 1.00/237& & \\
\\
PKS 0537-286 	& $\sim$2--40&7.5$^{+6.4}_{-5.5}$ & 1.46$^{+0.14}_{-0.13}$ & 3.09 & 0.74/197 & 1.85 & 0.63 &47.4& $<$139& $<$0.15$^h$\\
		&  &1.95 Fixed & 1.36$^{+0.06}_{-0.06}$ & 2.71 & 0.76/198& &\\
\\
S5 0836+71   	& $\sim$1--30&11.4$^{+3.0}_{-2.8}$ & 1.45$^{+0.05}_{-0.05}$ & 23.3  & 0.87/503 & 14.0 & 4.25&47.8 & $<$110& $<$0.11$^h$\\
		&  &2.78 Fixed & 1.32$^{+0.03}_{-0.03}$ & 19.3 & 0.97/504 & &\\
\\
PKS 1614+051$^i$ 	&$\sim$2--40 & 4.90 Fixed & 1.6 Fixed& 0.43& & 0.21$^i$&  0.10$^i$&  46.4$^i$& & \\
		&  &  &  & & \\
\\
PKS 2126-158 	&$\sim$2--40 & 13.6$^{+5.3}_{-4.3}$ & 1.68$^{+0.09}_{-0.09}$ & 13.1 & 1.01/331 & 5.51 & 2.20&47.9 & $<$107& $<$0.08\\
		&  &4.85 Fixed & 1.53$^{+0.04}_{-0.04}$ & 10.8 & 1.07/332 & &\\
\\
PKS 2149-306 	&$\sim$1--30 & 8.3$^{+1.8}_{-2.2}$ & 1.54$^{+0.05}_{-0.05}$ & 19.2 & 1.07/633 & 9.94 & 3.81 &47.8& $<$85$^h$& $<$0.17$^h$\\
		&  &1.91 Fixed & 1.42$^{+0.03}_{-0.02}$ & 16.3 & 1.13/634& &\\
\hline
\hline
\end{tabular}
\end{center}
\small
\pn
Note: Intervals are at 90\% confidence, for two interesting parameters when $N_{\rm H}$ is
free and for one interesting parameter when $N_{\rm H}$ is fixed at the Galactic value.
\pn
$ ^a$ Approximate energy range, in the quasar rest frame.
$ ^b$ SIS unabsorbed flux at 1 keV (observed frame) in units of
10$^{-4}$ photons cm$^{-2}$ s$^{-1}$ keV$^{-1}$.
$ ^c$ Absorbed flux between 2--10 keV (observed frame) in units of 10$^{-12}$ ergs cm$^{-2}$
s$^{-1}$, calculated with the SIS normalization only.
$ ^d$ Absorbed flux between 0.1--2 keV (observed frame) in units of 10$^{-12}$ ergs cm$^{-2}$
s$^{-1}$, extrapolated from the best-fit spectra (i.e. with the SIS normalization). 
$ ^e$ Intrinsic luminosity between 2--10 keV (quasar frame) in units of ergs s$^{-1}$,
calculated with the SIS only.
$ ^f$ 90\% limit on the equivalent width of the Fe K line measured in the observer frame, with
$\sigma$ = 0 eV and E = 6.4 keV (quasar frame), for one interesting parameter ($\Delta
\chi^2 = 2.71$). These values were about a factor of two worse if the line widths are as 
large as 0.5 keV.
$^g$ 90\% limit on the absorption edge depth with energy fixed at 7.1 keV (quasar frame).
$^h$ Value affected by the calibration uncertainties of the SIS between $\sim$ 1.8--2.4 keV.
An estimated conservative systematic error of 10 eV (observer frame) for the lines
and $\Delta \tau = 0.05$ for the edge depth have been added to these values.
$^i$ Quasar was detected but with too few counts to perform a spectral analysis. Fluxes
and luminosity are calculated assuming a photon index = 1.6 (i.e. the average value from the
sample) and Galactic absorption.

\vfill\eject
\centerline{TABLE 6: {\it ASCA} and {\it ROSAT} Combined Spectral Fits}
\ss
\centerline{A. Single Power Law ($N_{\rm H}$ free; $N_{\rm H}$ at z + $N_{\rm Hgal}$ )}
\begin{center}
\begin{tabular}{ccccc}
\hline
\hline
\multicolumn{1}{c}{Object}&
\multicolumn{1}{c}{Energy Range$^a$} &
\multicolumn{1}{c}{$N_{\rm H}$} &
\multicolumn{1}{c}{$\Gamma$} &
\multicolumn{1}{c}{$\chi^{2}_{red}$/dof} \\
\multicolumn{1}{c}{} &
\multicolumn{1}{c}{(keV)} &
\multicolumn{1}{c}{($10^{20}$ cm$^{-2}$)} &
\multicolumn{1}{c}{} &
\multicolumn{1}{c}{} \\
\hline
S5 0014+81    & $\sim$0.4--40 & 27.4$^{+4.0}_{-4.3}$& 1.71$^{+0.06}_{-0.07}$ & 1.03/572  \\
               &   & 554$^{+196}_{-170}$ & 1.70$^{+0.07}_{-0.06}$ & 1.03/572 \\
                 & \\
NRAO 140        &$\sim$0.2--20 & 31.1$^{+2.9}_{-2.7}$& 1.70$^{+0.04}_{-0.05}$ & 0.97/881  \\
                 &  &109$^{+20}_{-19}$ & 1.66$^{+0.05}_{-0.04}$ & 0.98/881 \\
                & \\
PKS 0438-436 	& $\sim$0.4--38&5.81$^{+2.66}_{-1.40}$ & 1.48$^{+0.10}_{-0.09}$ & 0.97/281  \\
	        &  &71.9$^{+44.1}_{-22.2}$ & 1.46$^{+0.09}_{-0.09}$ & 0.97/281 \\
	& \\
PKS 0537-286 	& $\sim$0.4--40&2.82$^{+0.97}_{-0.74}$ & 1.38$^{+0.08}_{-0.08}$ & 0.77/204  \\
                 &  &16.7$^{+19.3}_{-14.6}$ & 1.37$^{+0.08}_{-0.08}$ & 0.77/204\\
		& \\
PKS 2126-158 	& $\sim$0.4--40&10.4$^{+3.1}_{-2.3}$ & 1.63$^{+0.07}_{-0.08}$ & 0.99/411  \\
	        &  &116$^{+84}_{-51}$ & 1.58$^{+0.07}_{-0.06}$ & 1.01/411 \\
	&\\
\hline
\hline
\end{tabular}
\end{center}
\small
\pn
Note: Intervals are at 90\% confidence, for two interesting parameters.
\pn
$ ^a$ Approximate energy range, in the quasar rest frame.



\bs
\centerline{B. Complex models for the absorption}
\begin{center}
\begin{tabular}{ccccc|cccc}
\hline
\hline
\multicolumn{1}{c}{}&
\multicolumn{4}{c}{Warm Absorber (at z) + $N_{\rm Hgal}$} &
\multicolumn{4}{c}{Broken Power Law ($N_{\rm H} \equiv N_{\rm Hgal}$)}\\
\multicolumn{1}{c}{Object}&
\multicolumn{1}{c}{$N_{\rm W}$} &
\multicolumn{1}{c}{$\Gamma$} &
\multicolumn{1}{c}{$\xi$} &
\multicolumn{1}{c|}{$\chi^{2}_{red}$/dof} &
\multicolumn{1}{c}{$\Gamma_{\rm soft}$} &
\multicolumn{1}{c}{E$_{\rm break}$} &
\multicolumn{1}{c}{$\Gamma_{\rm hard}$} &
\multicolumn{1}{c}{$\chi^{2}_{red}$/dof} \\
\multicolumn{1}{c}{} &
\multicolumn{1}{c}{(cm$^{-2}$)} &
\multicolumn{1}{c}{} &
\multicolumn{1}{c}{(ergs cm$^{-2}$ s$^{-1}$)} &
\multicolumn{1}{c|}{} &
\multicolumn{1}{c}{} &
\multicolumn{1}{c}{(keV)} &
\multicolumn{1}{c}{} &
\multicolumn{1}{c}{} \\
\hline
S5 0014+81   & 741$^{+647}_{-298}$ & 1.72$^{+0.09}_{-0.07}$ & $<$ 191 & 1.03/571 	& -0.05$^{+0.70}_{-1.51}$ & 1.04$^{+0.16}_{-0.12}$ & 1.60$^{+0.04}_{-0.04}$ & 1.03/571 \\
		& \\
NRAO 140 & 305$^{+156}_{-124}$ & 1.72$^{+0.05}_{-0.06}$ & 93$^{+117}_{-76}$ & 0.98/880  & 0.85$^{+0.12}_{-0.13}$ & 1.57$^{+0.11}_{-0.10}$& 1.64$^{+0.05}_{-0.04}$ & 0.96/880   \\
		& \\
PKS 0438-436 & 210$^{+297}_{-158}$ & 1.49$^{+0.11}_{-0.11}$ & $<$ 120 & 0.97/280 & $<$0.5 & 0.68$^{+0.37}_{-0.33}$ & 1.41$^{+0.08}_{-0.09}$& 0.99/280 \\
         		& \\
PKS 0537-286 & 210$^{+789}_{-190}$ & 1.40$^{+0.12}_{-0.12}$ & $<$ 4350 & 0.77/203 & $<$1.32 & $<$3.8 & 1.47$^{+0.37}_{-0.16}$& 0.81/203 \\
       		& \\
PKS 2126-158 & 398$^{+552}_{-281}$ & 1.62$^{+0.08}_{-0.07}$ & $<$ 455 & 1.01/410&1.02$^{+0.20}_{-0.27}$ & 1.26$^{+0.42}_{-0.26}$ & 1.62$^{+0.11}_{-0.07}$& 1.00/410\\
         	&\\
\hline
\hline
\end{tabular}
\end{center}
\small
\pn
Note: Intervals are at 90\% confidence, for two interesting parameters.

\vfill\eject

\bs
\centerline{C. Complex emission models}
\begin{center}
\begin{tabular}{cccc|cccc}
\hline
\hline
\multicolumn{1}{c}{} &
\multicolumn{3}{c}{Bremsstrahlung model ($N_{\rm H}$ free)} &
\multicolumn{4}{c}{Reflection model ($N_{\rm H}$ free)} \\
\multicolumn{1}{c}{Object} &
\multicolumn{1}{c}{$N_{\rm H}$} &
\multicolumn{1}{c}{kT} &
\multicolumn{1}{c|}{$\chi^{2}_{red}$/dof} &
\multicolumn{1}{c}{$N_{\rm H}$} &
\multicolumn{1}{c}{$\Gamma$} &
\multicolumn{1}{c}{R} &
\multicolumn{1}{c}{$\chi^{2}_{red}$/dof} \\
\multicolumn{1}{c}{} &
\multicolumn{1}{c}{(cm$^{-2}$)} &
\multicolumn{1}{c}{keV} &
\multicolumn{1}{c|}{} &
\multicolumn{1}{c}{(cm$^{-2}$)} &
\multicolumn{1}{c}{} &
\multicolumn{1}{c}{(=A$_{\rm refl}$/A$_{\rm pl}$)} &
\multicolumn{1}{c}{} \\
\hline
S5 0014+81    	& 18.73$^{+3.71}_{-3.56}$ & 42.1$^{+8.00}_{-7.61}$ & 1.05/572     & 27.8 $^{+7.5}_{-5.7}$ & 1.75$^{+0.25}_{-0.10}$ & $<$ 0.70 & 1.04/571 \\
		& \\
NRAO 140        & 22.5$^{+2.1}_{-2.5}$ & 22.3$^{+2.6}_{-2.3}$  & 0.99/881          & 31.1$^{+3.4}_{-1.5}$ & 1.70$^{+0.11}_{-0.05}$ & $<$ 0.35 & 0.98/880 \\
		& \\
PKS 0438-436 	& 4.63$^{+1.68}_{-1.02}$ & 52.7$^{+30}_{-15.7}$ &
0.98/281     & 5.86$^{+6.21}_{-1.25}$ & 1.49$^{+0.26}_{-0.09}$ & $<$ 0.61 & 0.97/280 \\
		& \\
PKS 0537-286 	& 2.32$^{+0.83}_{-0.60}$ & 89.7$^{+75}_{-32}$ & 0.80/204          & 2.82$^{+1.00}_{-0.72}$ & 1.38$^{+0.15}_{-0.08}$ & $<$ 0.35 & 0.81/203 \\
		& \\
PKS 2126-158 	& 6.90$^{+1.53}_{-0.87}$ & 41.9$^{+10.0}_{-6.9}$ & 1.01/411       & 10.4$^{+3.9}_{-2.3}$ & 1.62$^{+0.26}_{-0.07}$ & $<$ 0.60 & 1.01/410 \\
		&\\
\hline
\hline
\end{tabular}
\end{center}
\small
\pn
Note: Intervals are at 90\% confidence, for two interesting parameters.


{\centerline{TABLE 7: Local Interstellar CO and IR Emission towards the Quasars}
\ss
\begin{center}
\begin{tabular}{ccccccccccc}
\hline
\hline
\multicolumn{1}{c}{} &
\multicolumn{1}{c}{} &
\multicolumn{1}{c}{} &
\multicolumn{1}{c}{CO$^{a}$} &
\multicolumn{1}{c}{} &
\multicolumn{1}{c}{IR$^{c}_{100{\mu}{\rm m}}$} &
\multicolumn{1}{c}{} \\
\multicolumn{1}{c}{Object}&
\multicolumn{1}{c}{$l$} &
\multicolumn{1}{c}{$b$} &
\multicolumn{1}{c}{Emission ?} &
\multicolumn{1}{c}{Ref.$^{b}$} &
\multicolumn{1}{c}{Emission ?} &
\multicolumn{1}{c}{Ref.$^{b}$}\\
\hline
S5 0014+81 	& 121.6113 & 18.8020 & possible & (1) & No & (6) \\
		&\\
PKS 0332-403 	& 244.7640 & -54.0749 & ... & ... & No & (6)\\
		&\\
NRAO 140        & 158.9997 & -18.7650 & Yes & (2)& No & (6)\\
                & \\
PKS 0438-436 	& 248.4109 & -41.5654 & ... & ... & No & (6)\\
		&\\ 
PKS 0537-286 	& 232.9400 & -27.2924  & ... & ... & No & (6)\\
		&\\
S5 0836+71   	& 143.5408 & 34.4257 & No & (3) & No & (6)\\
		&\\
PKS 2126-158 	& 35.9295  & -41.8679 & No & (4) & Yes & (6)\\
		&\\
PKS 2149-306 	& 17.0770 & -50.7845 & No & (5) & No & (6)\\
		&\\
\hline
\hline
\end{tabular}
\end{center}
\small
\pn
$ ^a$ An entry of ``Yes''/``No'' indicates $^{12}$CO J=1--0 line Galactic emission has/hasn't
been detected in the direction of the quasar (see \S 5.1.1.a).
\pn
$ ^b$ References: (1) Heithausen et al. (1993), (2) Bania et al. (1991), (3) Lisz \& Wilson
(1993), (4) Hartmann D., private communication, (5) Liszt (1994), (6) Moshir (1989)
\pn
$ ^c$ An entry of ``Yes''/``No'' indicates there is/isn't contamination by cirrus
(see \S 5.1.1.a).

\vfill\eject

\pn
{\bf Fig. 1}: $\chi^2$ contour plots in the $N_{\rm H}$--$\Gamma$ parameter space.
Full line contours represent 68\%, 90\% and 99\% confidence limits
obtained from {\it ROSAT} and {\it ASCA} (GIS+SIS).
The 90\% confidence contours obtained from fitting separately the GIS and SIS are
indicated in dashed lines. Best-fit values are indicated by a large
mark (+) for {\it ROSAT} and {\it ASCA} (GIS+SIS), and by a small mark ($\scriptstyle +$)
for GIS and SIS.
The vertical line represent the Galactic absorption obtained from radio
measurements at 21 cm with associated errors of 10$^{19}$ cm$^{-2}$ if taken from
Elvis et al. (1989) or with a conservative 30\% error, if taken from Dickey \& Lockman (1990).
\ms
{\bf Fig. 2}: Best-fit spectra and $\chi^2$ contour plots of the combined
{\it ROSAT} and {\it ASCA} data.
The data are fitted with a single absorbed power law model, and have been binned
to a S/N ratio higher than 4 for display purposes. Note that the
{\it ROSAT} counts (lower spectra) have been normalized by the PSPC geometric
area of 1141 cm$^2$. Therefore the {\it ROSAT}
counts should be read as ``normalized counts/sec/keV/cm$^2$''. The contours represent
the 68\%, 90\% and 99\% confidence limits, and the vertical lines are, like in Fig. 1,
the Galactic absorption and associated errors. For S5 0014+81, the energy of the Fe K
edge has been indicated.
\ms
{\bf Fig. 3}: X-ray photon indices vs. (a) redshift, (b) luminosity. Stars represent 
the {\it ASCA} slopes and errors between
$\sim$ 1--30 keV (quasar frame) for the 8 high-z RLQs (Table 5, fit with $N_{\rm H}$ free).
Triangles represent the {\it Ginga} slopes and errors between $\sim$ 2--20 keV for 18 low-z
RLQs as reported by Lawson \& Turner (1996). The horizontal shaded-line represents
the approximate value $<\Gamma> = 1.4$ of the $\sim$ 2--10 keV slope of the XRB (Gendreau et
al. 1995).
\ms
{\bf Fig. 4}: $\chi^2$ contour plots in the $N_{\rm H}$--$\Gamma$ parameter space
for 3C 273. Full line contours represent 68\%, 90\% and 99\%
confidence limits for SIS0+1. Dashed lines indicate the 90\% confidence
contours for SIS0 chip number 1 (S0c1) and SIS1 chip number 3 (S1c3),
fitted separately. Best-fit values are indicated with marks (+). The vertical
line represent the Galactic absorption (full line) and associated 30\% errors
(dotted line) from Dickey \& Lockman 1990.
\ms

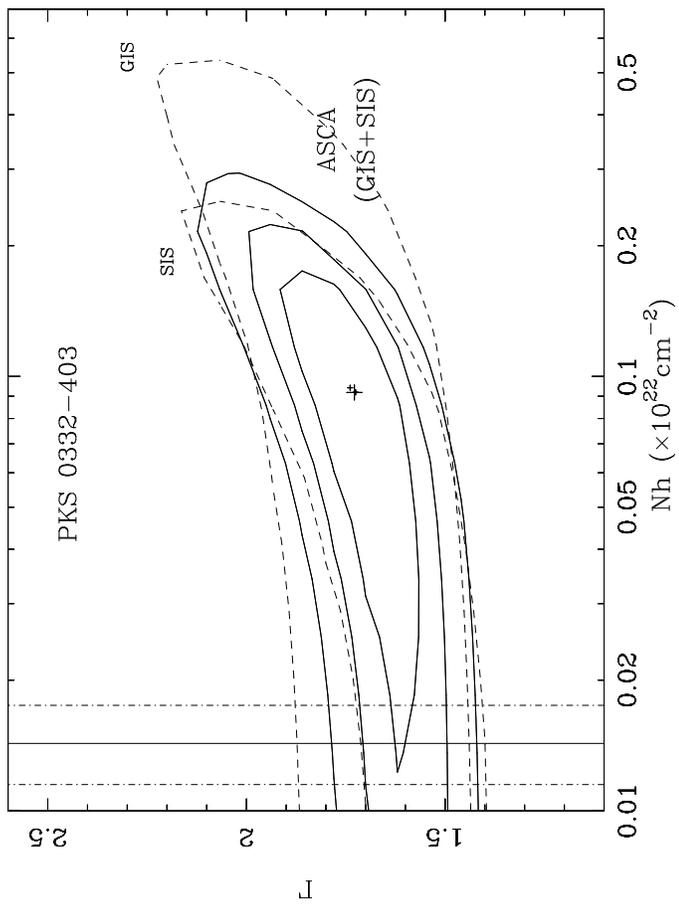
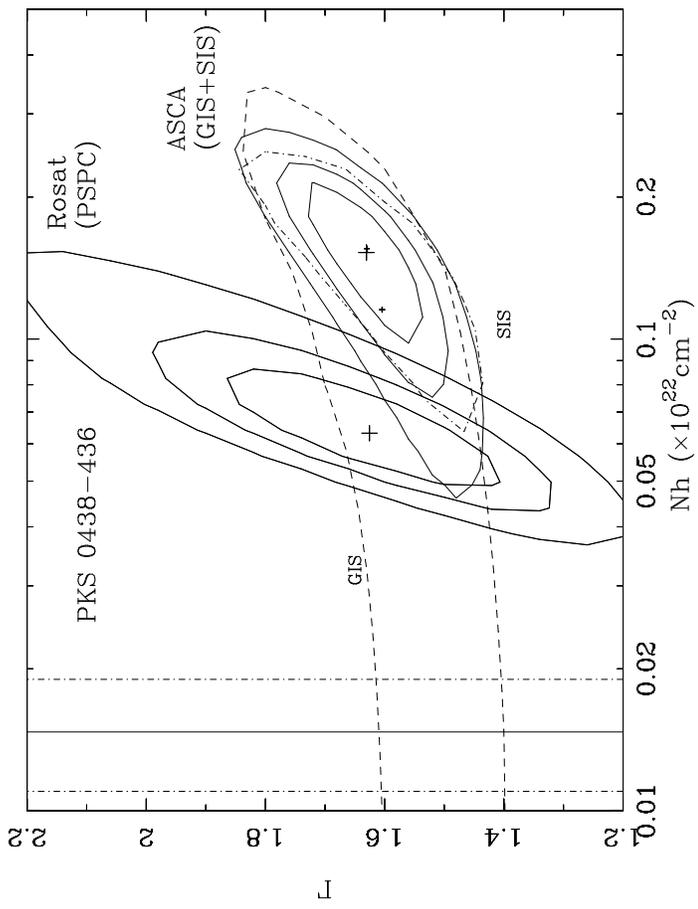
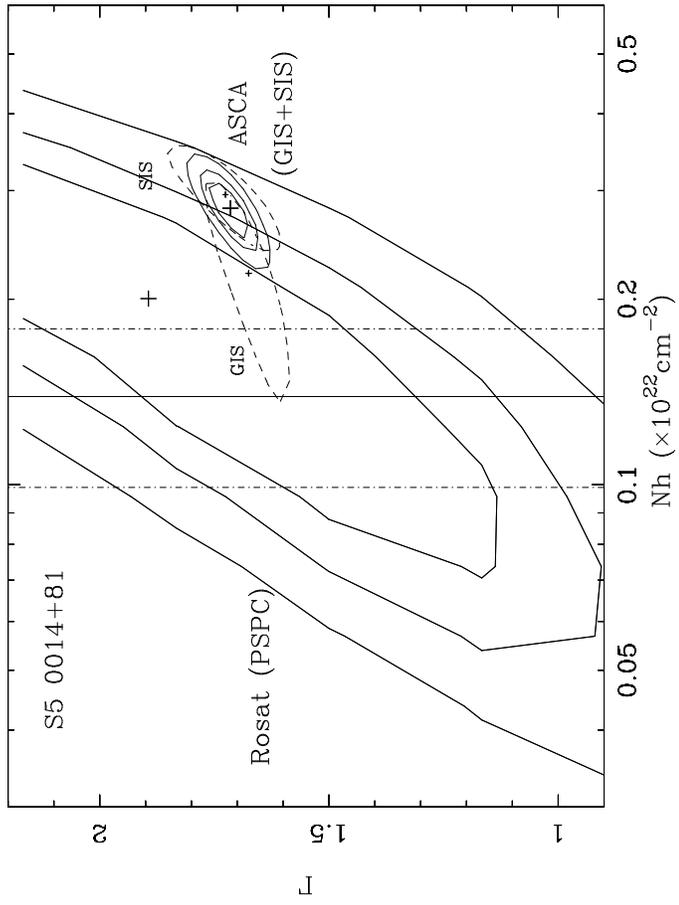
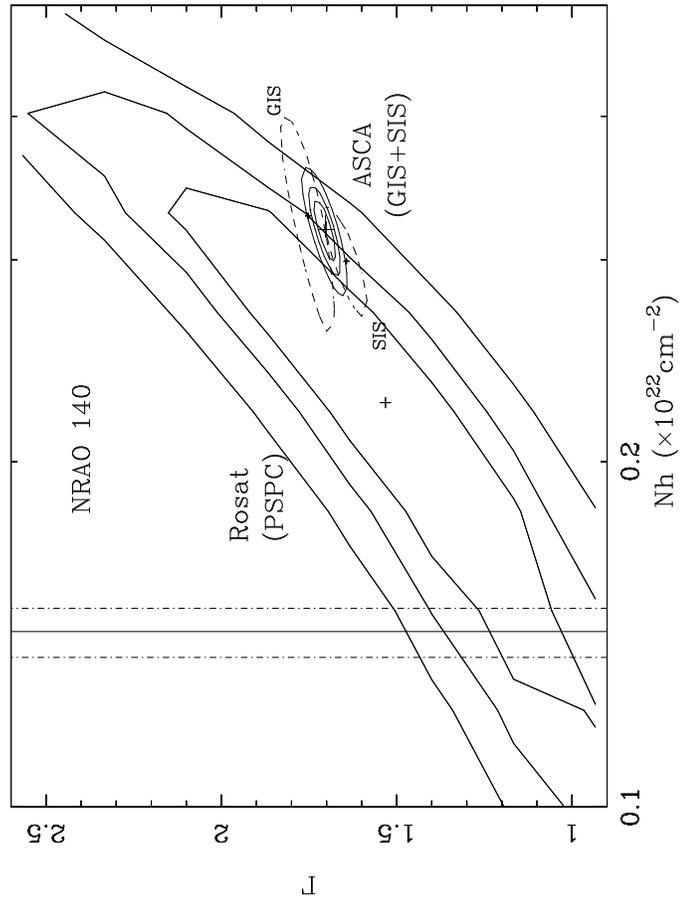
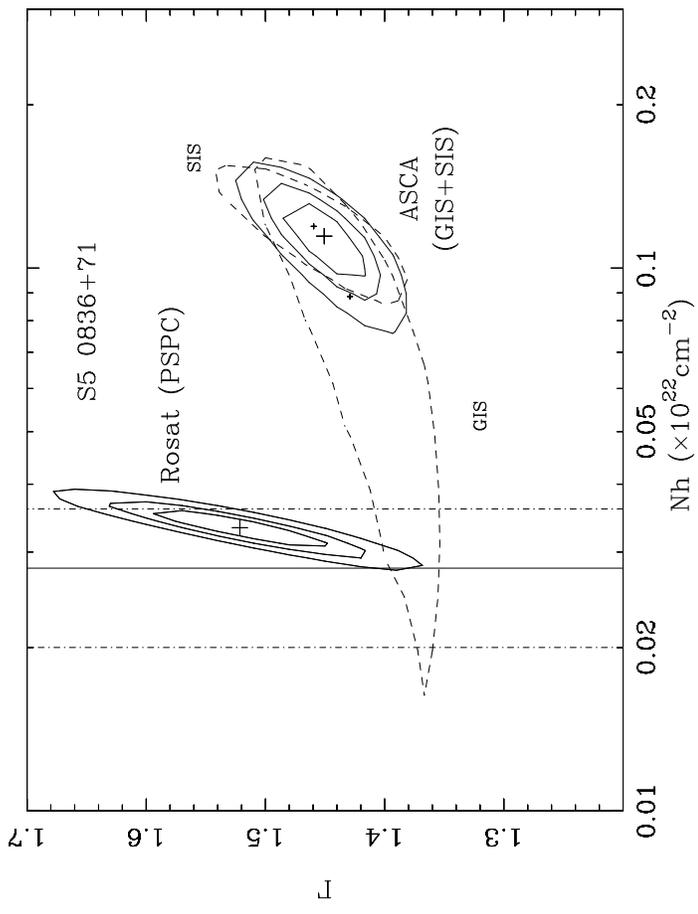
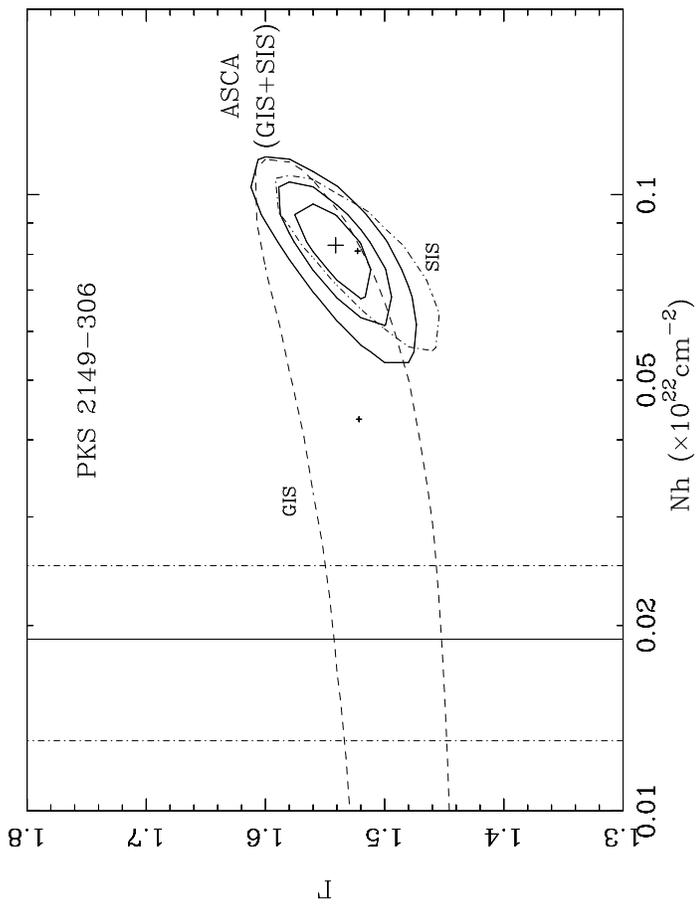
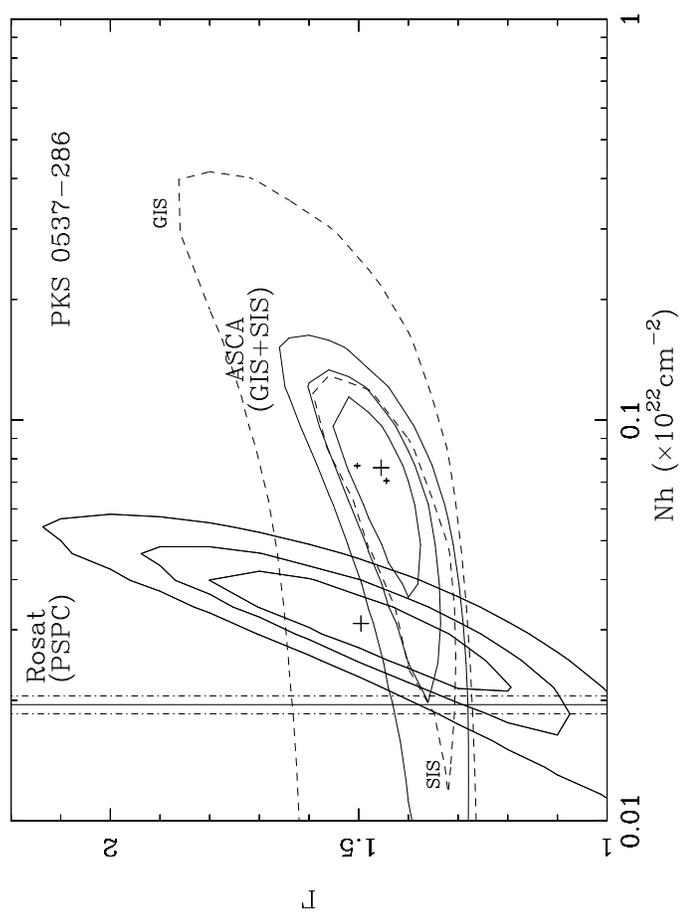
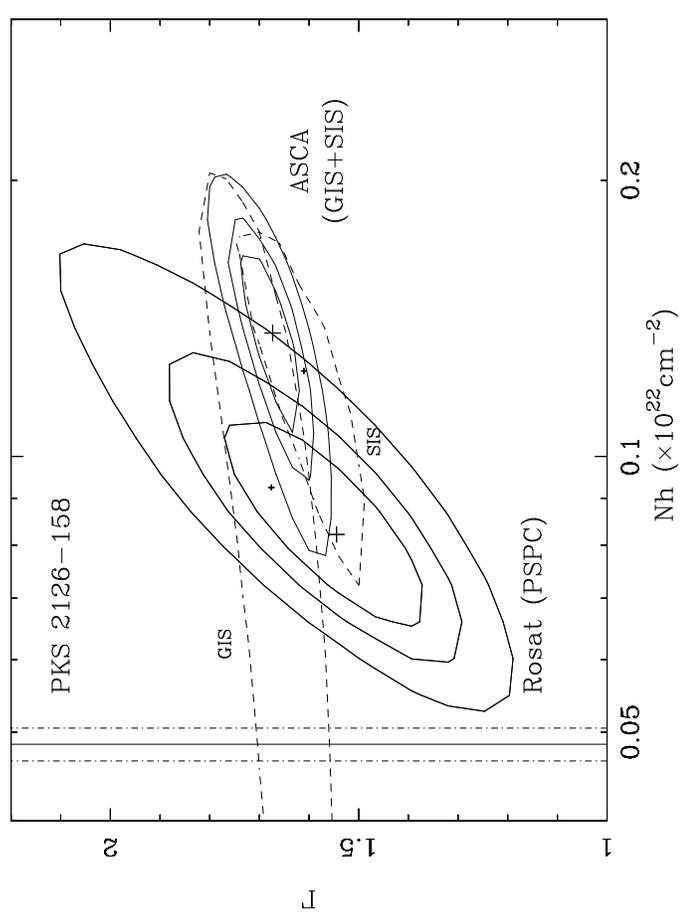
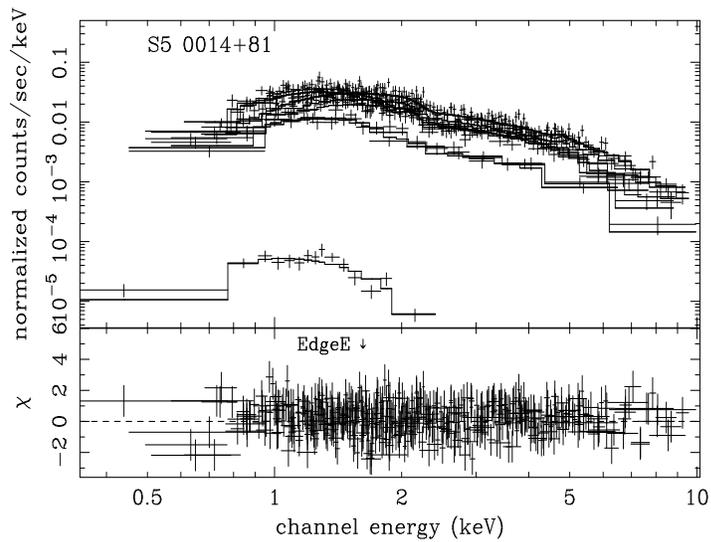
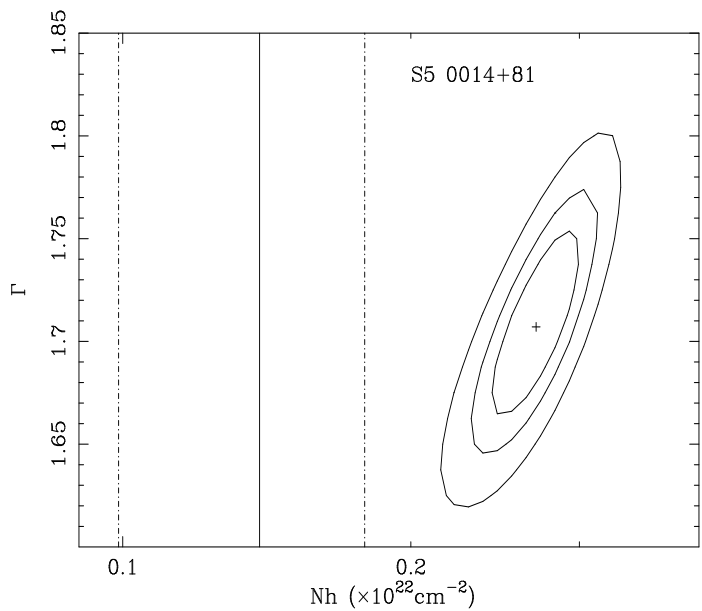
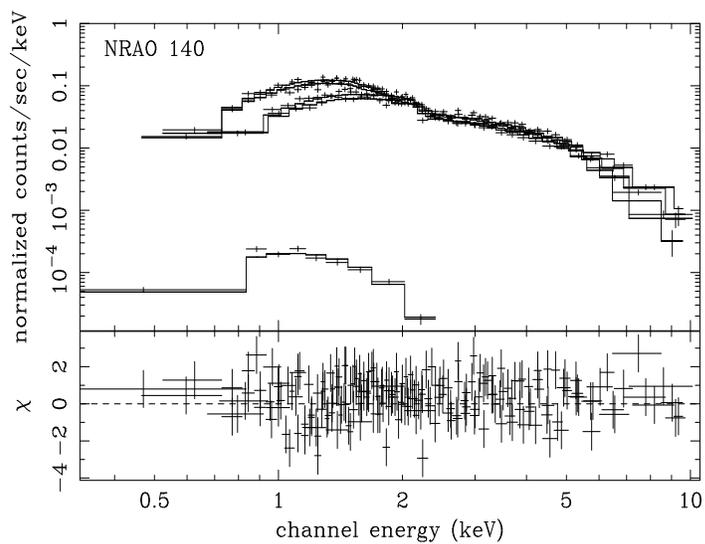
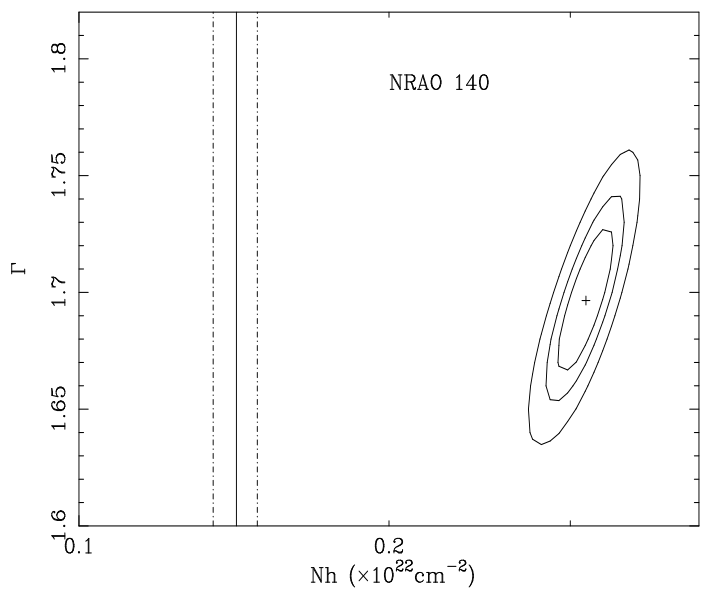
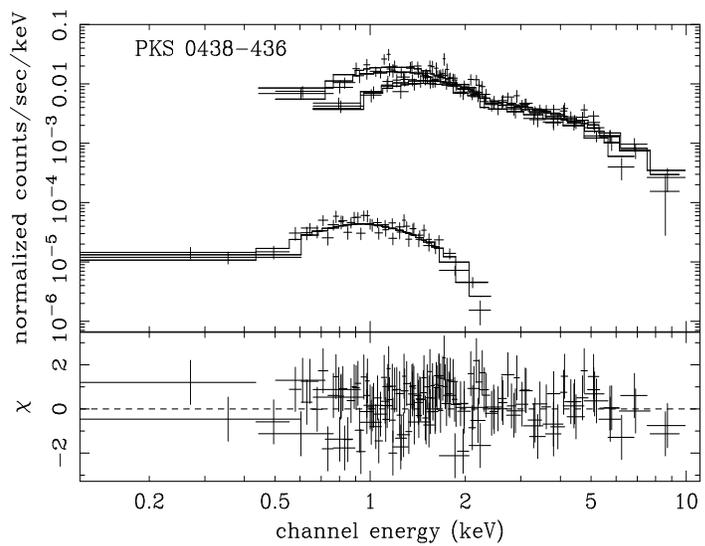
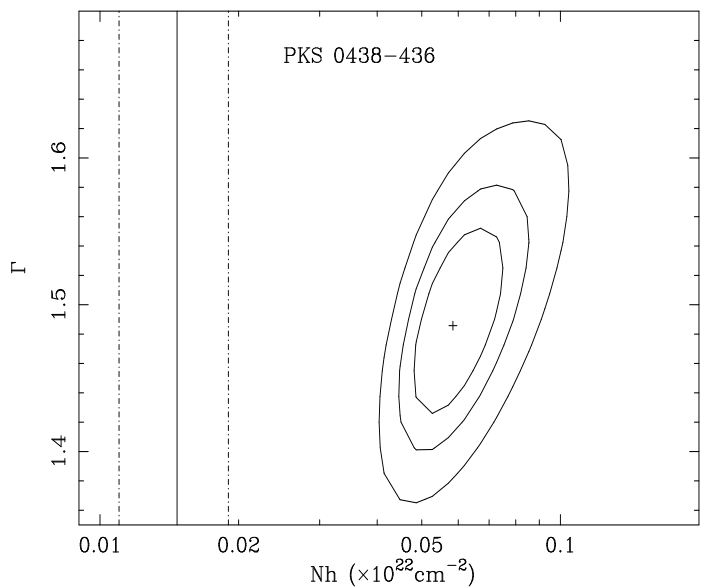
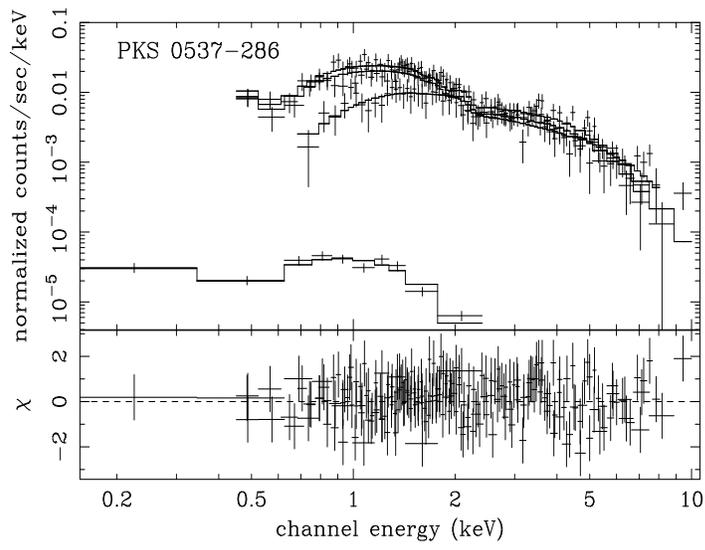
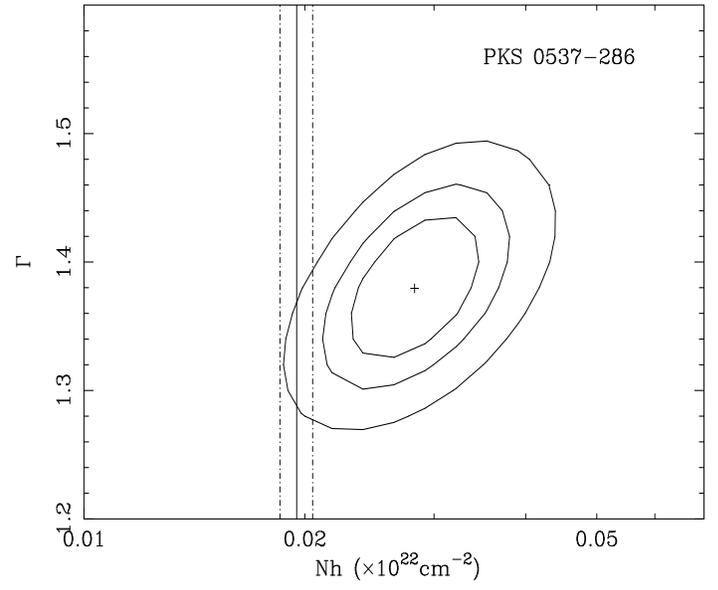
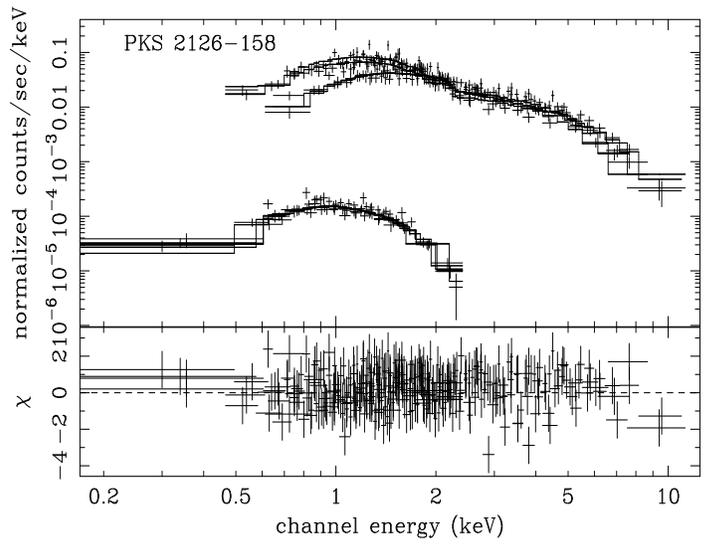
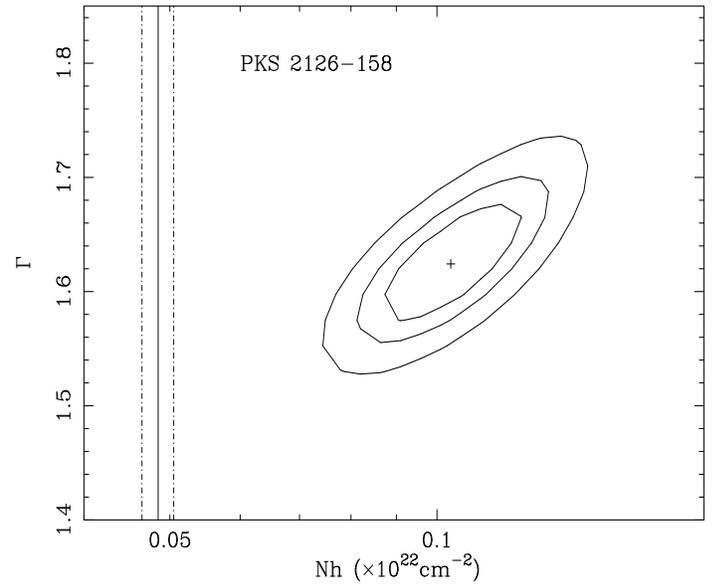
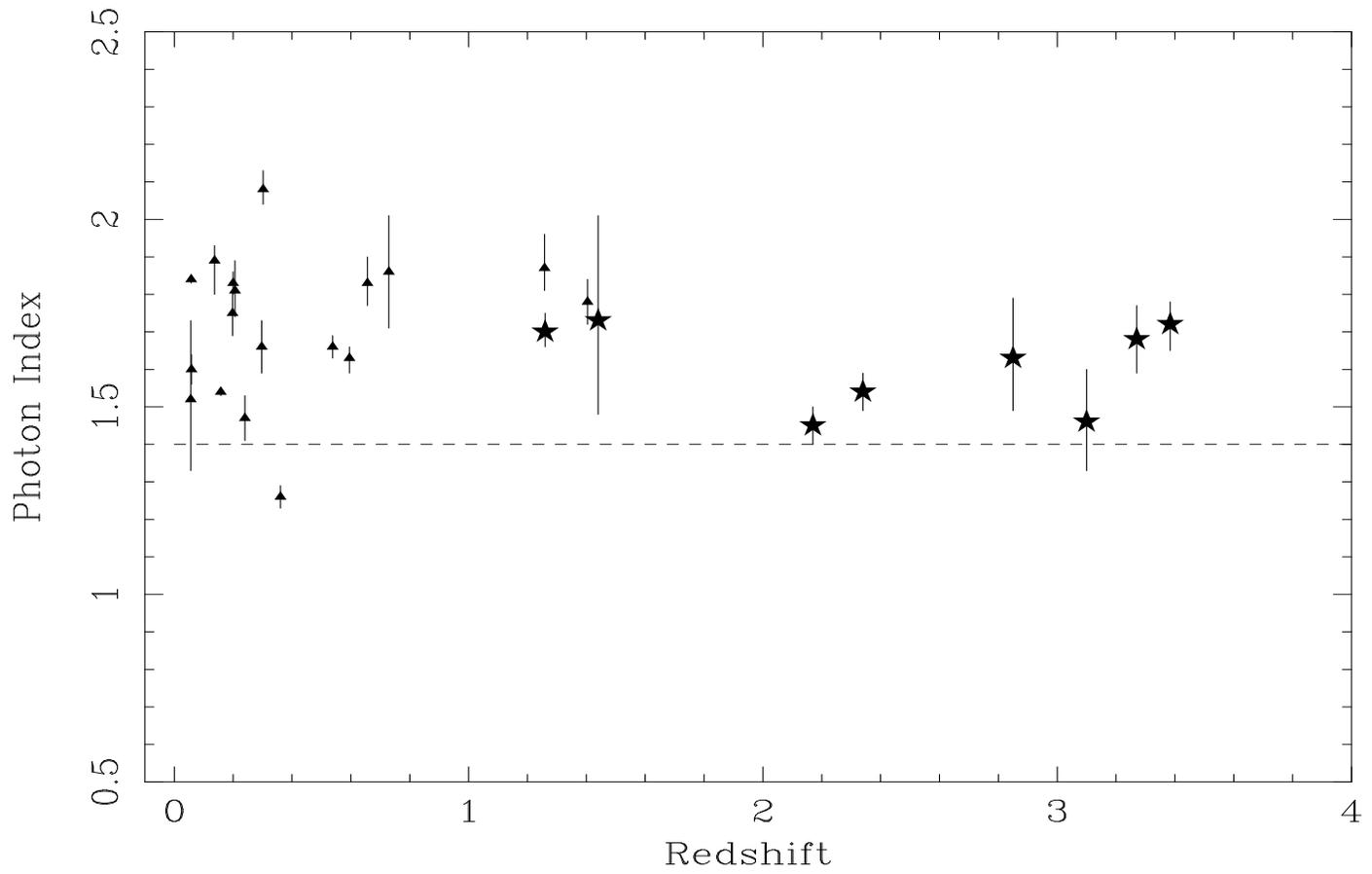
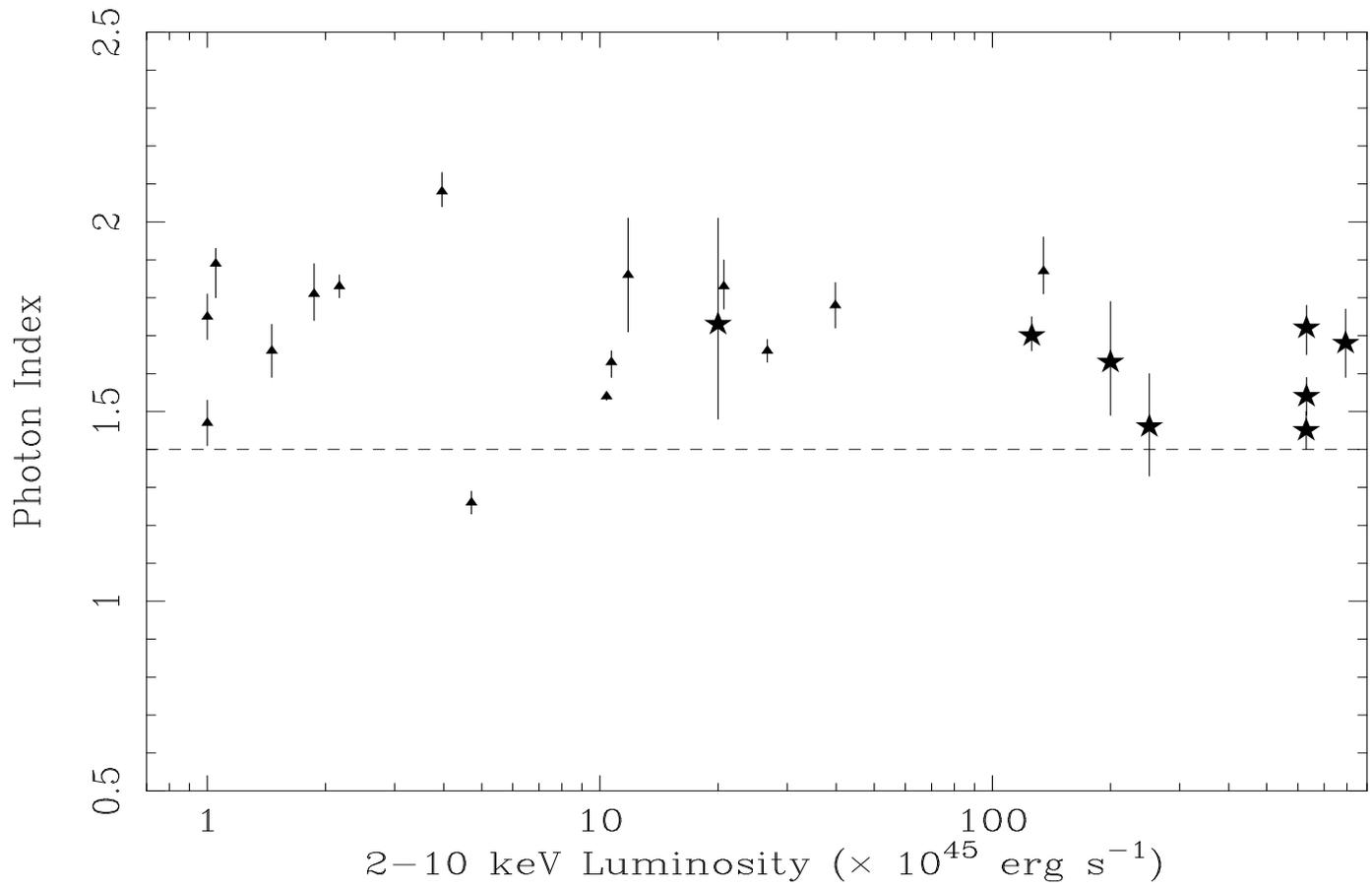
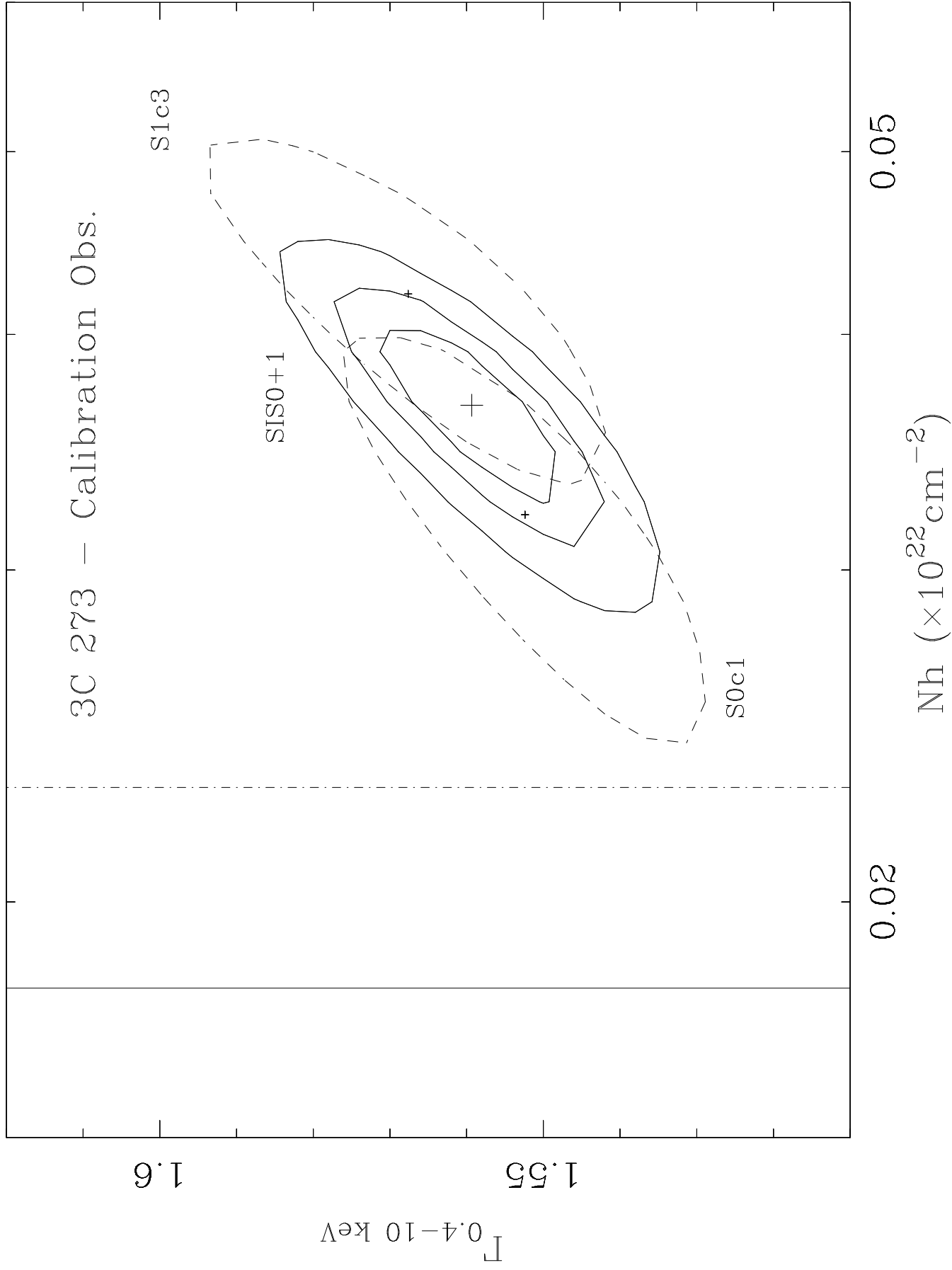
\end{document}